\documentclass[12pt]{iopart}  

\usepackage{dsfont}

\usepackage{graphics}
\usepackage{iopams}
\newcommand{\bra}[1]{\langle #1|}	% Bra- und Ketoperatoren werden defi2niert
\newcommand{\ket}[1]{|#1\rangle}
\newcommand{\braket}[2]{\langle #1|#2\rangle}
\begin{document}
  \title[Dissipative dynamics of a biased qubit coupled to a harmonic oscillator]{Dissipative dynamics of a biased qubit coupled to a harmonic oscillator: Analytical results beyond the rotating wave approximation}
  \author{Johannes Hausinger and Milena Grifoni}
  \address{Institut f\"ur Theoretische Physik, Universit\"at Regensburg, 93035 Regensburg, Germany}
  \ead{johannes.hausinger@physik.uni-regensburg.de}

\begin{abstract}
 We study the dissipative dynamics of a biased two-level system (TLS) coupled to a harmonic oscillator (HO), the latter interacting with an Ohmic environment. Using Van-Vleck perturbation theory and going to second order in the coupling between TLS and HO, we  show how the Hamiltonian of the TLS-HO system can be diagonalized analytically. Our model represents an improvement to the usually used Jaynes-Cummings Hamiltonian as an initial rotating wave approximation is avoided. By assuming a weak coupling to the thermal bath, analytical expressions for the time evolution of the populations of the TLS are found: the population is characterized by a multiplicity of damped oscillations together with a complex relaxation dynamics towards thermal equilibrium. The long time evolution is characterized by a single relaxation rate, which is largest at resonance and whose expression can be given in closed analytic form. 
\end{abstract}


\submitto{\NJP}


\maketitle

\section{Introduction}
In recent years the spin-boson model \cite{Leggett} -- \cite{GrifoniSpinBoson} has experienced a strong revival, as it is well suited to describe dissipative and decoherence effects on the dynamics of a two-level system (TLS) or qubit coupled to a bath. Crucial for the effects of the environment on the dynamics of the TLS is the shape of the spectral density of the harmonic bath. It is common to assume an Ohmic spectral density, which is linear in the continuous bath modes. In this work we concentrate on a so-called structured bath, for which the spectral density is Ohmic at low frequencies but exhibits a Lorenztian-shaped peak at a certain frequency $\Omega$. It has been shown in \cite{Garg} that a spin-boson model with such an effective spectral density can be exactly mapped  on the model of a TLS which is coupled to a single harmonic oscillator (HO) of frequency $\Omega$, where the latter feels the influence of an Ohmic bath.\newline
Due to its wide applicability the TLS-oscillator system has been object of intense research along the years. So it reflects for example the physics of single atoms with a large electric dipole moment coupled to the microwave photons of a cavity \cite{Raimond}, or quantum dots in photonic crystals \cite{Yoshie, Reithmaier}. More recently the model has received quite some attention in the field of quantum computation, where two-level systems are used to implement the two logical states of a qubit. We will especially focus on the solid-state implementation of such systems. Here, two prominent realizations of a qubit-oscillator system are the Cooper-pair box (CPB) \cite{Nakamura1999} -- \cite{Collin2004} coupled to a transmission line resonantor \cite{Blais} -- \cite{WallraffSideband} and the Josephson flux qubit \cite{Mooij1999} read out by a dc-SQUID \cite{vanderWal2000} -- \cite{Johansson2006}. Inspired by experiments with real atoms interacting with a cavity mode, one speaks for the CPB case of circuit quantum electrodynamics, as now the CPB plays the role of an artificial atom and the waveguide acts as a cavity. From such a setup one expects a huge step towards the realization of a quantum computer, as the transmission line resonator can be used to couple qubits together \cite{Makhlin, Blais2007}, store the information of qubits or to provide non-demolition read-out schemes \cite{Blais, Wallraff2005}. Concerning the flux qubit, the read-out usually happens through a damped dc-SQUID, which is inductively coupled to the qubit. However, through the SQUID enviromental noise is transferred to the qubit leading to decoherence and dissipation within its dynamics. The effect of this noise on the qubit depends very much on the strength $g$ of the coupling between qubit and SQUID and one faces a conflicting situation. On the one hand one wants a strong coupling for a good read-out resolution. On the other hand the coupling should be minimized to keep the negative effects of the environment as small as possible. In \cite{Tian, vanderWal} it has been shown that the qubit-SQUID system can be described by a spin-boson model with an effective spectral density $G_{\rm eff}(\omega)$ exhibting a peak at the plasma frequency $\Omega$ of the SQUID. Applying the above mentioned mapping an equivalent point of view is to consider the SQUID as an LC-circuit coupled to the Ohmic bath and model it as a harmonic oscillator. A detailed description of a nondestructive read-out scheme is e.g. given in \cite{Lupascu2004}. \newline
 The spin-boson model can be formally solved using e.g. real-time path integral methods \cite{Leggett, Weiss}. However, in order to get closed-form analytical results, approximations must be invoked. A quite common one is the so-called weak coupling approximation (WCA), which is perturbative in the bath spectral density \cite{Weiss}. However, it has been shown that for strong qubit-HO coupling $g$ and for small detuning $\delta = \Omega - \Delta_{\rm b}$, where $\Delta_{b}$ is the qubit energy splitting, such an approximation breaks down \cite{Thorwart}, as coherent exchange processes between TLS and oscillator are disregared. For an unbiased qubit the non-interacting blip approximation (NIBA) used in \cite{Wilhelm2004} --\cite{Nesi}  circumvents this problem as it is non-perturbative in the coupling $g$  and therefore takes correctly into account the influence of the oscillator on the  TLS. Moreover, it allows an analytic treatment of the dynamics. However, the NIBA is known to break down for a biased qubit at low temperatures \cite{Leggett, Weiss}.  Another approach, which treats the system non-perturbatively in the bath is the flow-equation renormalization method \cite{Kleff2003, Kleff2004}, where the spin-boson Hamiltonian is diagonalized using infinitesimal unitary transformations. However, whithin this approach analytical solutions are difficult to find. Recently a polaron transformation was used by Huang \etal to obtain analytically the population dynamics and confirm the Shiba's relation for an unbiased TLS \cite{Huang2008}.\newline
In the case in which the qubit and the HO are considered as the central quantum system being coupled to an Ohmic bath, the numerical, ab-initio quasiadiabatic propagator path-integral (QUAPI) method \cite{Makri1995_1, Makri1995_2} is a nice tool as it enables to cover both the resonant regime, where the oscillator frequency is close to the qubit energy splitting, and the dispersive regime with the oscillator being far detuned from the qubit \cite{Thorwart, Goorden2004, Goorden2005}. Moreover, it can be applied to a biased as well as to an unbiased TLS and therefore be used as a testbed for analytical results. For qubits being operated at the degeneracy point, which means an unbiased TLS, very often a rotating wave approximation (RWA) is applied \cite{Blais}, which is expected to be valid for small detuning and yields as starting point the Jaynes-Cummings Hamiltonian \cite{Jaynes1963, Shore1993}. This model was first used to study a two-state atom interacting with a single, close to resonance cavity mode of the electromagnetic field and predicts e.g. the repeated revival and collapse of Rabi oscillations within the atomic excitation probability. By condsidering the TLS-HO system in the representation of displaced HO states, Brito \etal were able to truncate the infinite Hilbert space of this system without loosing the effects of the HO on the TLS dynamics \cite{Brito2008}. However, so far none of these works could provide an analytical expression for the dynamics of the dissipative qubit being valid for zero as well as non-zero detuning and for both a biased and unbiased TLS.
In this work an analytic expression for the dissipative qubit's dynamics which includes the effects of a finite detuning and of a static bias is derived. Specifically, starting from the qubit-HO perspective, the eigenvalues and eigenfunctions of the non-dissipative TLS-HO system are found approximately using Van-Vleck perturbation theory up to second order in the coupling $g$. Notice that no rotating wave approximation is required. Dissipation effects are then evaluated by solving a Born-Markov master equation for the reduced density matrix in the system's eigenbasis. 
\newline The structure of the work is as follows. The dissipative TLS-HO Hamiltonian and the main dynamical quantities are introduced in \sref{TheModel}.  Inspired by the work of Goorden \etal \cite{Goorden2004, Goorden2005}, we demonstrate in \sref{NonDissTLSHO} how the eigenstates and eigenenergies of the non-dissipative Hamiltonian can be found approximately using Van-Vleck perturbation theory \cite{Shavitt1980, Cohen1992}. In this way we can provide an analytical formula for the non-dissipative dynamics whichs takes into account the full  Hilbert space of the qubit-HO system. After that, we show how for low temperatures ($k_{\rm B}T < \hbar \Omega, \hbar \Delta_{\rm b}$) this infinite Hilbert space can be truncated and discuss the relevant contributions of the HO to the dynamics. In \sref{InfluenceEnvironment} the influence of the environment is investigated, by looking at solutions of the Bloch-Redfield equations. Specifically, analytical expressions fo the TLS dynamics are obtained and compared with numerical solutions. The main physical features of the coupled TLS-HO system are discussed in \sref{DiscussionQubits}. To illustrate the effects of counter-rotating terms in the Hamiltonian of the qubit-HO system, which are neglected performing a RWA, we compare in \sref{CompJC} our calculations to results obtained from the Jaynes-Cummings model.

\section{The model} \label{TheModel}
In this section we introduce the Hamiltonian for a qubit coupled through a harmonic oscillator to a thermal bath. Further, a formula for the population difference between the qubit's two logical states is derived.

\subsection{The qubit-oscillator-bath system}
To set up the model we consider the Hamiltonian of a qubit-HO system, $\mathcal{H}_{\rm QHO}$, which is coupled to an environmental bath, $\mathcal {H}_{B}$, by the interaction Hamiltonian $\mathcal{H}_{\rm OB}$, so that the total Hamiltonian becomes
\begin{equation} \label{StartingHamiltonian}
  \mathcal{H} = \mathcal{H}_{\rm QHO} + \mathcal{H}_{\rm OB} + \mathcal {H}_{B}.
\end{equation} 
The Hamiltonian,
  $\mathcal{H}_{\rm QHO} = \mathcal {H}_{0} + \mathcal {H}_{\rm{Int}},$
consists of
\begin{equation}
   \mathcal {H}_0  = \mathcal {H}_{\rm TLS} + \mathcal {H}_{\rm HO} =- \frac{\hbar}{2} (\varepsilon \sigma_{\rm z} + \Delta_0 \sigma_{\rm x}) + \hbar \Omega B^\dagger B,
\end{equation}
 the Hamiltonian of the TLS/qubit and the harmonic oscillator, and the interaction term
\begin{equation}
  \mathcal {H}_{\rm{Int}} =   \hbar g \sigma_z (B^\dagger+B).
\end{equation} 
The Hamiltonian of the TLS is given in the subspace  $\{\ket{\rm L}, \ket{\rm R}\}$,  corresponding to  a clockwise or counterclockwise current in the superconducting loop of a three-junction Josephson qubit or more generally to the qubit's two logical states. In the case of a superconducting flux-qubit, the energy bias $\varepsilon$ can be tuned by an applied external flux, $\Phi_{\rm ext}$, and is zero at the so-called degeneracy point. The tunnelling amplitude is described by $\Delta_0$. For $\varepsilon \gg \Delta_0$ the states $\ket{\rm L}$ and $\ket{\rm R}$ are eigenstates of $\mathcal {H}_{\rm TLS}$, whereas at the degeneracy point those eigenstates are a symmetric and antisymmetric superposition of the two logical states. Further, $B$ and $B^\dagger$ are the annihilation and creation operator for the HO with frequency $\Omega$, and $g$ characterizes the coupling strength. We also introduce the energy splitting $\hbar \Delta_b \equiv \hbar \sqrt{\varepsilon^2 + \Delta_0^2}$ between the groundstate $\ket{\rm g}$ and the excited state $\ket{\rm e}$ of the TLS. Using the transformation
\begin{equation}
 R(\Theta) = \left( \begin{array}{cc}
 \cos \left(\Theta/
   2\right) & \sin
   \left(\Theta
   /2\right) \\
 -\sin \left(\Theta
  /2\right) & \cos
   \left(\Theta
   /2\right)
\end{array}\right)
\end{equation}
with $\tan \Theta  = - \Delta_0 / \varepsilon$ and $-\frac{\pi}{2} \leq \Theta < \frac{\pi}{2}$,
we obtain the Hamiltonian of the TLS in the this basis:
$\tilde \mathcal  { H}_{TLS} = R^T(\Theta) \mathcal {H}_{TLS} R(\Theta) = -\frac{\hbar \Delta_b}{2} \tilde \sigma_{\rm z}$. 
 The states $\ket{\rm R}$ and $\ket{\rm L}$ become in the energy basis
\begin{eqnarray}
    \ket{R} &= \cos (\Theta/2) \ket{\rm g} +\sin (\Theta/2) \ket{\rm e} , \label{LocaltoEn1}\\
   \ket{L} &= -\sin (\Theta/2) \ket{\rm g} +\cos (\Theta/2) \ket{\rm e}.  \label{LocaltoEn2}
\end{eqnarray}
The Hamiltonian $\mathcal {H}_{\rm HO}$ is diagonal in the eigenbasis $\{\ket{j}\}$ with $j=0,\ldots,\infty$ being the occupation number: $\mathcal {H}_{\rm HO} =\sum_j \hbar j \Omega \ket{j}\bra{j}$. 
For the eigenbasis of the combined Hamiltonian  $ \tilde \mathcal {H}_0 \equiv \tilde \mathcal{H}_{\rm TLS} + \mathcal{H}_{\rm HO}$ we write
\begin{equation} \label{eigenbasis}
  \{ \ket{j} \otimes \ket{\rm g}; \ket{j} \otimes \ket{\rm e} \} \equiv \{ \ket{j {\rm g}} ; \ket{j {\rm e}}  \}.
\end{equation} 
Following Caldeira and Leggett \cite{CaldeiraLeggett}, we model the environmental influences originating from the circuitry surrounding the qubit and the oscillator as a bath of  harmonic oscillators being coupled bilinearly to the HO. Thus, the environment is described by
  $\mathcal{H}_{\rm B} = \sum_k \hbar \omega_k b^\dagger_k b_k$
and the interaction Hamiltonian is
\begin{equation} \label{CouplingHam}
  \mathcal{H}_{\rm OB} = (B^\dagger + B) \sum_k \hbar \nu_k (b_k^\dagger + b_k) +  (B^\dagger + B)^2 \sum_k \hbar \frac{\nu_k^2}{\omega_k}.
\end{equation}
The operators $b_k^\dagger$ and $b_k$ are the creation and destruction operator, respectively, for the $k^{th}$ bath oscillator, $\omega_k$ is its frequency and $\nu_k$ gives the coupling strength.  The whole bath can be described by its spectral density, which we consider to be Ohmic:
\begin{equation} \label{OhmicSpecDens}
  G_{ \rm{Ohm}} (\omega) = \sum_k  \nu_k^2 \delta (\omega - \omega_k) = \kappa \omega. 
\end{equation} 
 In \cite{Garg} it is shown that the above model is equivalent to that of a TLS being coupled directly to a harmonic bath including the single oscillator of frequency $\Omega$; i.e., a spin-boson model \cite{Leggett, Weiss}  with a peaked effective spectral density,
\begin{equation} \label{EffDens}
  G_{\rm eff}= \frac{2 \alpha \omega \Omega^4}{(\Omega^2 - \omega^2)^2 + (2 \pi \kappa \omega \Omega)^2}.
\end{equation} 
The relation between $\alpha$ and the coupling parameter $g$ between the qubit and the HO is
  $g = \Omega \sqrt{\alpha / (8 \kappa)}$ \cite{Tian, vanderWal}.
This second perspective is suitable for calculating the dynamics of the qubit using a path-integral approach, as it was done for example in \cite{Nesi} for the case of an unbiased qubit ($\varepsilon = 0$). The approach in \cite{Nesi}, however, being based on the NIBA \cite{Weiss}, is not suitable to investigate the low temperature dynamics of a biased TLS. Thus, in this manuscript we will consider  the TLS and the single oscillator as central quantum system and solve the Bloch-Redfield master equations for the density matrix of this system, which are valid  also for the case of a biased TLS.

\subsection{The population difference} \label{PopulationDifference}
The main goal of this work is to determine the dynamics $P(t)$ of the qubit. That means, we wish to calculate the population difference 
\begin{equation} \label{dynamics}
  P(t) = \Tr_{\rm TLS} \{ \sigma_{ \rm z} \rho_{\rm red} (t) \} = \bra{\rm R}  \rho_{\rm red}(t) \ket{\rm R} - \bra{\rm L}  \rho_{\rm red}(t) \ket{\rm L}
\end{equation}
between the $\ket{R}$ and $\ket{L}$ states of the qubit.
The reduced density matrix of the TLS,
\begin{equation}
  \rho_{\rm red}(t) = \Tr_{\rm{HO}} \{  \rho (t)\} = \Tr_{\rm{HO}} \Tr_{\rm B} \{ W(t) \}
\end{equation} 
is found after tracing out the oscillator and bath degrees of freedom from the total density matrix
  $W (t) = \rme^{- \frac{\rmi}{\hbar} \mathcal {H} t} W (0) \rme^{\frac{\rmi}{\hbar}  \mathcal {H} t}$. In turn $\rho(t) = \Tr_{\rm B} \{ W(t) \}$ is
the reduced density matrix of the qubit-HO system. How to calculate this density matrix will be shown later.  After some algebra, illustrated in more detail in \ref{AppPt},  we arrive at an expression for $P(t)$, given in terms of diagonal and off-diagonal elements of $\rho(t)$ in the TLS-HO eigenbasis $\{\ket{n} \}$. It reads
\begin{equation} \label{PtQubitHO}
  P(t) = \sum_n p_{nn}(t) + \sum_{\stackrel{n,m}{n > m}} p_{nm}(t)
\end{equation}  
where
\numparts
\begin{eqnarray}
  \fl  p_{nn}(t) = \sum_j \left\{ \cos \Theta \biggl[ \braket{j {\rm g}}{n}^2 - \braket{j {\rm e}}{n}^2 \biggr]  + 2 \sin \Theta \braket{j {\rm g}}{n} \braket{j {\rm e}}{n} \right\} \rho_{n n}(t) \label{p0},\\
  \fl p_{nm}(t) = 2 \sum_j \biggl\{ \cos \Theta \biggl[\braket{j {\rm g}}{n} \braket{m}{j {\rm g}} - \braket{j {\rm e}}{n} \braket{m}{j {\rm e}}    \biggr] \nonumber\\
     + \sin \Theta \biggl[\braket{j {\rm e}}{n} \braket{m}{j {\rm g}} + \braket{j {\rm e}}{m} \braket{n}{j {\rm g}}    \biggr] \biggr\} \Re \{ \rho_{nm}(t) \} \label{pnm}
\end{eqnarray}
\endnumparts
with $\rho_{nm}(t) = \bra{n} \rho(t) \ket{m}$.  How to determine the eigenstates of $\mathcal{H}_{\rm QHO}$ is described in the next section.

\section{Energy spectrum and dynamics of the non-dissipative TLS-HO system} \label{NonDissTLSHO}
In this section we show how to find  the eigenvalues of the unperturbed qubit-HO Hamiltonian $\mathcal{H}_{\rm QHO}$ approximately by using Van-Vleck perturbation theory \cite{Shavitt1980, Cohen1992}. The idea is to take advantage of the degenerate or doublet structure of the energy spectrum of the uncoupled ($g=0$) TLS-HO system near resonance, e.g. at $\Delta_{\rm b} \approx \Omega$. Then, as long as the perturbation is small compared to the energy separation of the different doublets, the full Hamiltonian will exhibit a similar spectrum of bundled energy levels. 

\subsection{Energy spectrum} \label{SecEnSpec}
\begin{figure}[h]
\begin{flushright}
  \resizebox{.9\linewidth}{!}{
  \includegraphics{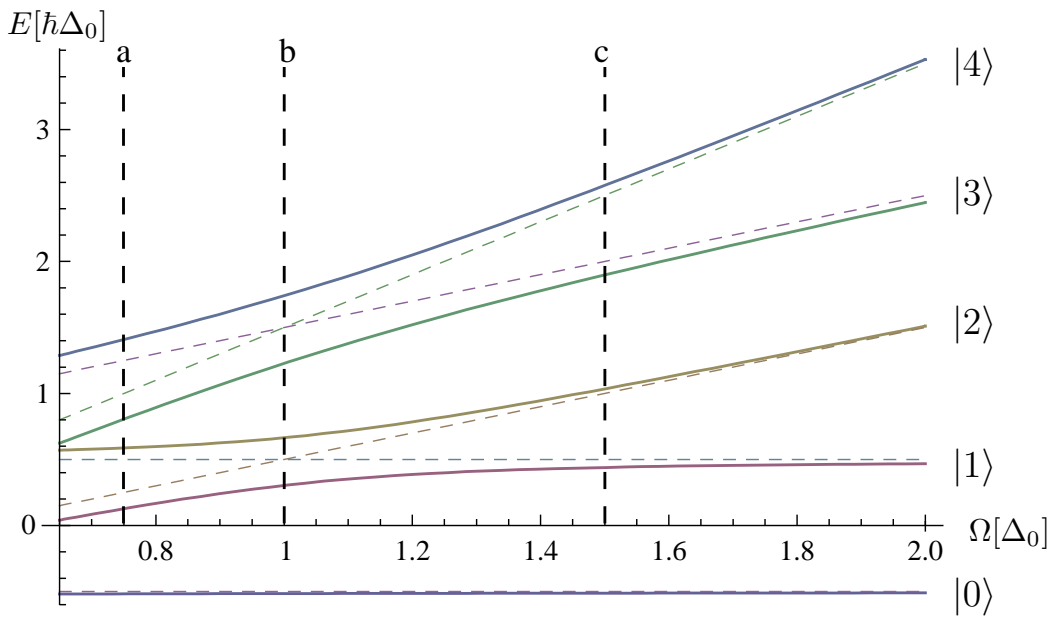}}
\end{flushright}
\caption{Energy spectrum of the coupled TLS-HO system vs. the oscillator frequency $\Omega$. Solid lines show the energy levels for the five lowest energy states  with the coupling being switched on ($g=0.18$) and the TLS being unbiased ($\varepsilon=0$). Frequencies and energies are given in units of $\Delta_0$ and $\hbar \Delta_0$, respectively. For comparison the energy levels for the uncoupled case are also given (dashed lines). At resonance ($\Omega = \Delta_{\rm b}$) the spectrum exhibits avoided crossings, whereas it approaches the uncoupled case away from resonance. The vertical dashed lines visualize three different situations: the negatively detuned regime (line a), the resonant case (line b) and the positively detuned regime (line c).  \label{EnSpecCoup}}
\end{figure}

 The eigenenergies of the uncoupled TLS-HO system are immediately found by applying the Hamiltonian $\tilde \mathcal { H}_0 =\tilde \mathcal { H}_{\rm TLS}+\mathcal { H}_{\rm HO}$ on the eigenstates in  \eref{eigenbasis}:
\begin{equation} \label{EigenEnNoCoup}
 \fl \tilde \mathcal { H}_0 \ket{j {\rm g}}  = \left( -\frac{\hbar \Delta_b}{2} + \hbar j \Omega \right) \ket{j {\rm g}} \quad {\rm and} \quad
   \tilde \mathcal { H}_0 \ket{j {\rm e}} = \left( \frac{\hbar \Delta_b}{2} + \hbar j \Omega \right) \ket{j {\rm e}}.
\end{equation}
The dashed lines in \fref{EnSpecCoup} show the energy spectrum corresponding to \eref{EigenEnNoCoup} vs. the oscillator frequency $\Omega$  for the five lowest eigenstates.
Except for the groundstate, $\ket{0 {\rm g}}$, the  states $\ket{(j+1) {\rm g}}$ and $\ket{j {\rm e}}$ are degenerate in the resonant case $(\Omega = \Delta_b)$. Close to  resonance the spectrum exhibits a doublet structure. 
With the coupling being switched on, the full Hamiltonian  $\mathcal {H}_{\rm QHO}$ reads
\begin{eqnarray} \label{HQHOEn}
 \fl \tilde \mathcal { H}_{\rm QHO} \equiv R^\dagger \mathcal {H}_{\rm QHO} R =\tilde \mathcal { H}_0 +\tilde  \mathcal { H}_{\rm Int} \nonumber\\
      = -\frac{\hbar \Delta_b}{2} \tilde \sigma_{\rm z} + \hbar \Omega B^\dagger B + \hbar g \left(  \frac{\varepsilon}{\Delta_b}\tilde \sigma_{\rm z}- \frac{\Delta_0}{\Delta_b}\tilde \sigma_{\rm x} \right) (B+B^\dagger)
\end{eqnarray}
in the basis  $\{ \ket{j {\rm g}} ; \ket{j {\rm e}}  \}$.  
In order to diagonalize the Hamiltonian $ \tilde \mathcal { H}_{\rm QHO}$  we consider $ \tilde \mathcal {H}_{\rm Int}$ as a small perturbation, which is resonable as long as $g \ll \Delta_b, \Omega$.
Applying Van-Vleck perturbation theory we construct an effective Hamiltonian, 
\begin{equation} \label{VanVleckTrafo}
 \tilde \mathcal { H}_{\rm eff} = \rme^{\rmi S} \tilde \mathcal {H}_{\rm QHO} \rme^{-\rmi S}, 
\end{equation} 
having the same eigenvalues as $\tilde \mathcal {H}_{\rm QHO}$ but no matrix elements connecting states which are far off from degeneracy. Thus, $\tilde \mathcal { H}_{\rm eff}$ will be block-diagonal with all quasi-degenerate energy levels being in one common block. As in our case always two states are nearly degenerate, each block of $\tilde \mathcal {H}_{\rm eff}$ builds a two-by-two matrix. This can be easily diagonalized in order to determine the eigenstates.
Following \cite{Shavitt1980, Cohen1992} we calculate the transformation matrix $S$ up to second order in $g$. The general formulas for both an arbitrary Hamiltonian and $\tilde \mathcal {H}_{\rm QHO}$ are given in \ref{AppVanVleck}.
The only surviving matrix elements of the effective Hamiltonian, apart from the ones being of zeroth order in $g$, are 
\begin{equation} 
  \left(\tilde \mathcal{H}_{\rm eff}\right)^{(1)}_{j {\rm e};{(j+1)\rm g}} = \left(\tilde \mathcal{ H}_{{\rm eff}}\right)^{(1)}_{(j+1){\rm g};j{\rm e}} = \hbar \Delta \sqrt{j+1} \textrm{\quad with \quad} \Delta = -\frac{g \Delta_0 }{\Delta_b} \label{Heff1},
\end{equation}
and 
\begin{eqnarray}
   \left(\tilde \mathcal{H}_{\rm eff}\right)^{(2)}_{j{\rm e};j{\rm e}} &= - \frac{\hbar  \varepsilon^2}{\Delta_b^2 \Omega}g^2 + j \frac{\hbar \Delta_0^2}{\Delta_b^2 (\Delta_b+\Omega)} g^2\equiv \hbar (W_1 - j W_0), \\
   \left(\tilde \mathcal{H}_{\rm eff}\right)^{(2)}_{j{\rm g};j{\rm g}} &= \hbar [W_1 + (j+1) W_0] . \label{Heff2}
\end{eqnarray}
Thus,  $\tilde \mathcal{ H}_{\rm eff}=\tilde \mathcal{ H}^{(0)}_{\rm eff} + \tilde \mathcal{ H}^{(1)}_{\rm eff} + \tilde \mathcal{ H}^{(2)}_{\rm eff}$ has the matrix structure
\begin{eqnarray} \label{HeffMatrix}
\fl \tilde \mathcal{ H}_{\rm eff} = \hbar \left(\begin{array}{ c|cc|c}
                                   			    \ddots &  &  & \\ \hline
								  & \frac{\Delta_b}{2}+j \Omega + W_1 - j W_0 & \sqrt{j+1} \Delta &  \\
								& & &\\
								 & \sqrt{j+1} \Delta & -\frac{\Delta_b}{2} +(j+1) \Omega + W_1 +(j+2) W_0 & \\ \hline
								 & & & \ddots
                                                        \end{array} \right),\nonumber\\
\end{eqnarray}
where the section shown corresponds to the basis states $ \ket{j{\rm e}}$ and $\ket{(j+1){\rm g}}$. From this form it is easy to calculate the eigenstates and eigenenergies.
The groundstate $\ket{0}_{\rm eff} \equiv \ket{0{\rm g}}$, which is an eigenstate of $\tilde \mathcal{ H}_{\rm eff}$, has the eigenenergy
\begin{equation} \label{GroundstateEn}
  E_0 = \hbar \left( -\frac{\Delta_b}{2}+W_0+W_1 \right). 
\end{equation} 
The other eigenstates of $\tilde \mathcal{ H}_{\rm eff}$ are, $j \geq 0$,
\numparts 
\begin{eqnarray} 
  \ket{2j+1}_{\rm eff} =\cos \left( \frac{\alpha_j}{2}  \right) \ket{(j+1){\rm g}} + \sin \left( \frac{\alpha_j}{2} \right) \ket{j {\rm e}}, \label{VanVleckEigenstate1} \\
  \ket{2j+2}_{\rm eff} = - \sin \left( \frac{\alpha_j}{2}  \right) \ket{(j+1){\rm g}} + \cos \left( \frac{\alpha_j}{2} \right) \ket{j {\rm e}} \label{VanVleckEigenstate2},
\end{eqnarray}
\endnumparts
corresponding to the eigenenergies 
\begin{equation} \label{Eigenenergies}
\eqalign{
\fl E_{2j+1/2j+2} =  \hbar \left[ (j+\frac{1}{2}) \Omega + W_1 + W_0 \mp \frac{\delta_j}{2 \cos \alpha_j} \right]\\
                             = \hbar \left[ (j+\frac{1}{2}) \Omega + W_1 + W_0 \mp  \frac{1}{2} \sqrt{\delta_j^2+4(j+1) |\Delta|^2} \right], }
\end{equation}
\begin{equation} \label{deltaj}
\fl {\rm with \quad}  \delta_j = \Delta_b - \Omega - 2 (j+1) W_0, \textrm{\quad} \tan \alpha_j = \frac{2 \sqrt{j+1} |\Delta|}{\delta_j}  \textrm{\quad and \quad} 0 \leq \alpha_j < \pi.
\end{equation} 
By construction these are also eigenenergies of $\tilde \mathcal{H}_{\rm QHO}$.
 Using the transformation (\ref{VanVleckTrafo}) we get the eigenvectors of $\tilde \mathcal{H}_{\rm QHO}$ as
\begin{equation} \label{QubitSQUIDEigenstates}
  \fl \ket{0} = \rme^{-\rmi S} \ket{0}_{\rm eff},  \textrm{\quad}   \ket{2j+1} = \rme^{-\rmi S} \ket{2j+1}_{\rm eff} \textrm{\quad and \quad} \ket{2j+2} = \rme^{-\rmi S} \ket{2j+2}_{\rm eff}.
\end{equation}
The energy spectrum of $\tilde \mathcal{H}_{\rm QHO}$ is shown in \fref{EnSpecCoup} for the case of an unbiased TLS ($\varepsilon =0$). We want to emphasize that our findings are also valid for the more general case $\varepsilon \neq 0$. At resonance, where the spectrum for the uncoupled case is degenerated, avoided crossings can be seen. The gap between two formerly degenerated levels for $\Omega = \Delta_b$ is 
\begin{equation} \label{DistanceQuasiEn}
  E_{2j+2} - E_{2j+1} = 2 \hbar \sqrt{j+1} g + \Or (g^3), 
\end{equation}
which is as predicted by the Jaynes-Cumming model \cite{Jaynes1963, Shore1993}. 
As we will show in \sref{CompJC},  the second order correction $W_0$ in \eref{HeffMatrix}, whichleads to a shift in the resonance frequency,  is a result of the counter-rotating terms in $\tilde \mathcal{H}_{\rm QHO}$. As such it can be interpreted as a Bloch-Siegert shift \cite{BlochSigert}.

\subsection{Dynamics of the qubit for the non-dissipative case} \label{DynamicsNoDiss}
With the coupling to the bath being turned off, the time evolution of the density matrix of the qubit-HO system is given by   $\rho (t) = \rme^{- \frac{\rmi}{\hbar} \tilde  \mathcal {H}_{\rm QHO} t} \rho (0) \rme^{\frac{\rmi}{\hbar} \tilde \mathcal {H}_{\rm QHO} t}$ and consequently 
\begin{equation} \label{DensMatNoDiss}
   \rho_{nm}(t) = \bra{n} \rho(t) \ket{m} = \rme^{- \rmi \omega_{nm} t} \rho_{nm}(0)
\end{equation} 
with $\omega_{nm} = \frac{1}{\hbar} (E_n - E_m)$. With that \eref{PtQubitHO} becomes
\begin{equation} \label{PtQubitSQUID}
  P(t) = p_0 + \sum_{\stackrel{n,m}{n > m}} p_{nm}(0) \cos{\omega_{nm} t},
\end{equation}  
where we defined $p_0 \equiv \sum_n p_{nn}(0)$.
From \eref{PtQubitSQUID} we notice that the dynamics of the qubit is characterized by an infinite number of oscillation frequencies rather than showing Rabi oscillations with a single distinct frequency. This is clearly a consequence of the coupling of the HO to the TLS.
Further we assume that at $t=0$ the qubit starts in the state $\ket{R}$ and that the occupation numbers of the HO are Boltzmann distributed, so that
\begin{equation}
  \rho(0) = \ket{\rm R}\bra{\rm R} \frac{1}{\rm Z} e^{-\beta \mathcal{ H}_{\rm HO}},
\end{equation} 
where ${\rm Z} = \rme^{\hbar \beta \Omega/2} / ( 1- e^{-\beta \hbar \Omega})$
 is the partition function of the oscillator and $\beta = ({\rm k_B} T)^{-1}$ denotes the inverse temperature of the system. In the TLS-HO eigenbasis this becomes
\begin{equation} \label{StartingConditions}
  \eqalign{
  \fl \rho_{nm}(0) = \bra{n} \rho(0) \ket{m}= \frac{1}{\rm Z} \sum_{j=0}^\infty \rme^{-\hbar \beta \Omega (j+\frac{1}{2})} \left[ \cos \left( \frac{\Theta}{2} \right) \braket{n}{j{\rm g}} + \sin \left( \frac{\Theta}{2} \right) \braket{n}{j{\rm e}} \right]\\
              \times  \left[ \cos \left( \frac{\Theta}{2} \right) \braket{j{\rm g}}{m} + \sin \left( \frac{\Theta}{2} \right) \braket{j {\rm e}}{m} \right].}
\end{equation} 

\subsection{Low temperature approximation}\label{LowTempApprox}
With \eref{PtQubitSQUID} we found a formula which describes using the approximate eigenenergies and eigenstates in \eref{Eigenenergies} and \eref{QubitSQUIDEigenstates}  the non-dissipative dynamics up to second order in $g$, thereby taking into account all oscillator levels. Thus, we still have to deal with an infinite Hilbert space. Typically experiments, see e.g. in \cite{Wallraff, Chiorescu2004}, run in a temperature regime for which $\beta^{-1} \lesssim \hbar \Omega, \hbar \Delta_{\rm b}$.
Considering the exponential function in \eref{StartingConditions} we assume the higher oscillator levels to be only sparsely populated and the maximum value of the sum in \eref{StartingConditions} is truncated to  $j=1$.
Nevertheless, states $\ket{j {\rm g/e}}$ with $j>1$ still play a role in the dynamics. In fact, due to the Van-Vleck transformation $\exp(-\rmi S)$, for example the state
\begin{equation}
  \ket{8} = \rme^{-\rmi S} \ket{8}_{\rm eff} = \rme^{-\rmi S} \left [-\sin \left( \frac{\alpha_3}{2}\right) \ket{4 {\rm g}} + \cos \left( \frac{\alpha_3}{2}\right) \ket{3 {\rm e}} \right] 
\end{equation} 
yields nonvanishing contributions to the matrix elements $\braket{n}{1 {\rm g}}$ and $\braket{n}{1 \rme}$ occurring in \eref{StartingConditions} due to the fact that the  energy eigenstates \eref{QubitSQUIDEigenstates} of the coupled TLS-HO system are made of linear combinations which involve also these states. \newline
  Using \eref{pnm} together with \eref{DensMatNoDiss} one finds that coefficients $p_{nm}(0)$ with $n \geq 7$ are of higher than second order in $g$. The same is valid for $p_{50}$, $p_{60}$, $p_{55}$ and $p_{66}$. Thus, those terms play no role in our calculation of $P(t)$. Furthermore, $\rme^{-\frac{3}{2} \beta \Omega} (g/\Delta_{\rm b} \Omega)^2 \ll 1.$
Neglecting also these contributions we find that $p_{n, m} \ll 1$ for $n \geq 5$.
In the end it will be sufficient to concentrate on eigenstates of $\tilde \mathcal{H}_{\rm QHO}$ up to $\ket{4}$. This trunctation leaves us with ten possible oscillation frequencies $\omega_{nm}$, where $n,m = 0,1,\ldots,4$ and $n>m$.

\begin{figure}[h]
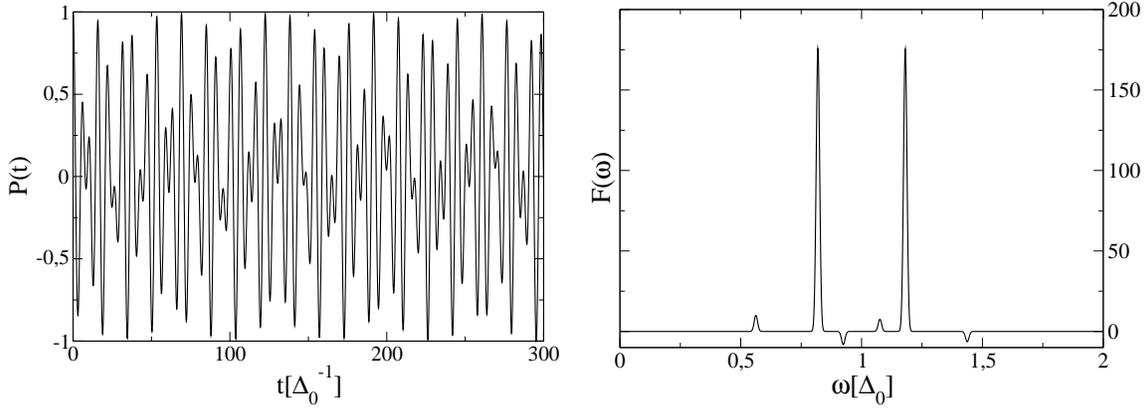

\begin{flushright}
\resizebox{.47 \linewidth}{!}{\includegraphics{Figures/figure2a.eps}}
\hspace{.1cm}
\resizebox{.47\linewidth}{!}{\includegraphics{Figures/figure2b.eps}}
\end{flushright}
\caption{Left-hand graph: Dynamics of the population difference $P(t)$ for the unbiased TLS-HO system at resonance ($\Omega = \Delta_0$) with $g=0.18$ and $\beta = 10$. Right-hand graph: Fourier transform $F(\omega)$ of $P(t)$ for the unbiased system. The peaks are situated around $\omega = \Delta_0$ according to \eref{Omega10NoBias} - \eref{Omega42NoBias}. Clearly, $\omega_{10}$ and $\omega_{20}$ are the dominating frequencies. In order to visualize the delta-functions, finite widths have been artificially introduced.  \label{PtNoBiasRes}}
\end{figure}

As an example we calculate the dynamics of an unbiased TLS ($\varepsilon =0$). Here the coefficients $p_0$, $p_{30}(0)$, $p_{40}(0)$, $p_{21}(0)$ and $p_{43}(0)$ vanish due to symmetry, so that
\begin{equation}
  \eqalign{
  \fl P(t) = p_{10} \cos \left( \omega_{10} t\right) + p_{20} \cos \left( \omega_{20} t\right) + p_{31} \cos \left( \omega_{31} t\right) + p_{41} \cos \left( \omega_{41} t\right)\\
     + \, p_{32} \cos \left( \omega_{32} t\right) + p_{42} \cos \left( \omega_{42} t\right).}
\end{equation} 
Additionally as a benchmark we consider the mostly studied resonant case, where $\Omega =\Delta_b = \Delta_0$. In this case we find with \eref{Eigenenergies} the transitions frequencies
\numparts
\begin{eqnarray}
  \omega_{10} = \Delta_0 - g, \quad   \omega_{20} = \Delta_0 + g, \label{Omega10NoBias}\\
  \omega_{31} = \Delta_0 + (1-\sqrt{2}) g, \quad   \omega_{41} = \Delta_0 + (1+\sqrt{2}) g,\\
  \omega_{32} = \Delta_0 - (1+\sqrt{2}) g, \quad \omega_{42} = \Delta_0 - (1-\sqrt{2}) g. \label{Omega42NoBias}
\end{eqnarray} 
\endnumparts
The  dynamical quantity $P(t)$ and its Fourier transform are shown in \fref{PtNoBiasRes}. One clearly sees the influence of the coupling to the HO on the dynamics of the TLS. Instead of Rabi oscillations with a single frequency, $P(t)$ oscillates with six different frequencies, which are in the Fourier spectrum symmetrically located around the point $\omega = \Delta_0$. Among those frequencies $\omega_{10}$ and $\omega_{20}$ are dominating. They correspond to transitions between the first or second energy level of the qubit-HO system and its groundstate and their weight is almost equal. To summarize, one notices that due to the coupling with the oscillator additional frequencies are induced into the qubit dynamics. Theoretically, the number of those frequencies is infinite. At low temperatures, however, transitions between the lower energy levels of the system are clearly dominating. Again, for simplicity we have shown here the case of an unbiased TLS. For $\varepsilon \neq 0$ the behaviour is similar only that in the Fourier spectrum the weight difference of the two dominating peaks will be more pronounced. 

\section{The influence of the environment} \label{InfluenceEnvironment}
In the preceding section we neglected the influence of the bath on the qubit-HO system. Yet, in order to model a realistic situation, we have to pay attention to environmental influences, as they lead to decoherence and dissipation in the dynamics of the qubit, which is harmful for quantum computing application. Thus, we will now consider the full Hamiltonian $\mathcal{H}$.

\subsection{Master equation for the qubit-HO system} \label{SecMasterEqu}
As shown in \sref{PopulationDifference}, we need for the calculation of the population difference  $P(t)$ the density matrix $\rho(t)$ of the qubit-HO system. Starting from the Liouville equation of motion for the full density matrix $W(t)$ of $\mathcal{H}$,
\begin{equation} \label{Liouville}
  \rmi \hbar \frac{\partial W(t)_{\rm I}}{\partial t} = \left[ \mathcal{H}_{OB}(t)_{\rm I}, W(t)_{\rm I} \right],
\end{equation}
where the index  stands for the interaction picture and  following \cite{Blum, Louisell}, we can provide a Born-Markov master equation for $\rho(t)$ being in the Schr\"odinger picture and expressed in the basis of the eigenstates of $\tilde \mathcal{H}_{\rm QHO}$:
\begin{equation} \label{MasterEqQOBasis}
 \dot \rho_{nm}(t) = -\rmi \omega_{nm} \rho_{nm} (t) + \pi \sum_{kl} \mathcal{L}_{nm,kl} \rho_{k,l}(t).
\end{equation}
The free dynamics of the system is given by the first term of the right-hand side in the above equation. The rate coefficients are defined as
\begin{equation} \label{RateCoefficients}
  \eqalign{
  \fl \mathcal{L}_{nm,kl} = \left[ G(\omega_{nk}) N_{nk} - G(\omega_{lm}) N_{ml}  \right] X_{nk} X_{lm} \\
         -\delta_{ml} \sum_{l^\prime} G(\omega_{l^\prime k}) N_{l^\prime k} X_{n l^\prime} X_{l^\prime k}  + \delta_{nk} \sum_{k^\prime} G(\omega_{l k^\prime}) N_{k^\prime l} X_{l k^\prime} X_{k^\prime m}}
\end{equation}
with
$N_{nm} = \frac{1}{2} [ \coth (\hbar \beta \omega_{nm}/2) - 1]$  and $X_{nm} = \bra{n} \left( B^\dagger + B  \right) \ket{m}$.
For the derivation of the master equation, besides the Born and Markov approximations, some more assumptions have been made, which we briefly mention. First, we consider our system and the bath to be initially (at $t=0$) uncorrelated; i.e., 
  $W(0)_{\rm I} = \rho(0) \rho_{\rm B}(0)$
with
  $\rho_{ \rm B}(0) = Z_{\rm B}^{-1} \rme^{-\beta \mathcal{H}_{\rm B}} $
and $Z_{\rm B}$ the partition function of the bath. Further, with the bath consisting of infinite degrees of freedom, we assume the effects of the interaction with the qubit-HO system to dissipate away quickly, such that the bath remains in thermal equilibrium for all times $t$:
  $W(t)_{\rm I} = \rho(t)_{\rm I} \rho_{\rm B} (0).$ 
Besides, an initial slip term which occurs due to the sudden coupling of the system to the bath is neglected \cite{Weiss}. And as last approximation the Lamb-shift of the oscillation frequencies $\omega_{nm}$ was not taken into account \cite{Louisell}.

\subsection{Matrix elements} \label{OscillatorMatrixElements}
In \eref{RateCoefficients} $X_{nm}$ describes matrix elements of the operator $X=(B + B^\dagger)$ in the qubit-HO eigenbasis. By use of \eref{VanVleckEigenstate1}, \eref{VanVleckEigenstate2} and \eref{QubitSQUIDEigenstates} those states were  expressed in the basis $\{\ket{j {\rm g}}; \ket{j {\rm e}} \}$, and we will also calculate the oscillator matrix elements in this basis. For that purpose the operator $\tilde X = \rme^{\rmi S} \left(B^\dagger + B  \right) \rme^{-\rmi S}$ is defined. Four different situations can be distinguished. There are matrix elements were neither the oscillator nor the qubit state are changed, namely
  $ \bra{j {\rm g}} \tilde X \ket{j {\rm g}} = - 2  L_0$  and  $\bra{j {\rm e}} \tilde X \ket{j {\rm e}} =  2  L_0$ with  $L_0 = \varepsilon g / \Delta_{\rm b} \Omega$.
We see that those elements are independent of $j$, the occupation number of the oscillator. Next, we look at the case where a single quantum is emitted or absorbed from the oscillator and get
 $\bra{j {\rm g}} \tilde X \ket{(j+1) {\rm g}} = \sqrt{j + 1} (1 +  L_{\rm osc})$ and $ \bra{j {\rm e}} \tilde  X \ket{(j+1) {\rm e}} =  \sqrt{j + 1} (1 - L_{\rm osc})$  
with
\begin{equation}
L_{\rm osc}  = \frac{(2 \Delta_b + 3 \Omega) \Delta_0^2}{\Delta_b^2 \Omega (\Delta_b + \Omega)^2}g^2.
\end{equation} 
For a transition within the qubit we have
  $ \bra{j  {\rm g}} \tilde X \ket{j {\rm e}} = \Delta_0 g / \Delta_b (\Delta_b + \Omega) \equiv  L_{\rm q}$.
And finally, if the qubit and the oscillator state are changed simultaneously, one obtains
$\bra{j {\rm g}} \tilde X \ket{(j+1) {\rm e}} = \sqrt{j + 1}  L_{\rm q,osc}^+$ and  $\bra{j {\rm e}} \tilde X \ket{(j+1) {\rm g}} =  \sqrt{j + 1}  L_{\rm q,osc}^-$, 
\begin{equation}
 \fl {\rm where}  \quad L_{\rm q,osc}^+ = \frac{4 \varepsilon \Delta_0}{ \Delta_b^2 (\Delta_b + \Omega) (\Delta_b + 2 \Omega)} g^2 \quad {\rm and} \quad L_{\rm q,osc}^- = \frac{ - 4 \varepsilon \Delta_0}{ \Delta_b^2  \Omega (\Delta_b - 2 \Omega)}g^2.
\end{equation}   
Comparing the magnitude of the transition terms, we notice that those consisting in changes of the oscillator occupation only are the dominant ones, as they have a part which is of zeroth order in $g$. Further, for the case in which the qubit is operated at the degeneracy point $ L_0$ and $ L_{\rm q,osc}^{+/-}$ vanish. With those results we can calculate the matrix elements $X_{nm}$. They are given in  \ref{AppOscillatorMatrixElements}.

\subsection{Dynamics in the dissipative case}\label{DynamicsDiss}
Like in \sref{LowTempApprox} we assume the system to be operated at low temperatures and thus take as highest qubit-HO state the eigenstate $\ket{4}$. For determination of $P(t)$ the formulas of \sref{PopulationDifference} can be used. Unlike in the non-dissipative case $\rho(t)$ is not given anymore by the simple expression \eref{DensMatNoDiss}. Rather we have to solve a system of coupled differential equations, namely \eref{MasterEqQOBasis}. To do this analytically we will follow three different approaches and compare them finally to the numerical solution of \eref{MasterEqQOBasis}. We start by introducing
\begin{equation} \label{AnalyAnsatz}
  \rho_{nm} (t) = \rme^{-\rmi \omega_{nm} t} \sigma_{nm}(t),
\end{equation}
which yields the set of differential equations for $\sigma_{nm}:$
\begin{equation} \label{MasterWithAnsatz}
  \dot \sigma_{nm} (t) = \pi \sum_{k l} \mathcal{L}_{nm,kl} \rme^{\rmi (\omega_{nm} - \omega_{kl})t} \sigma_{kl}(t).
\end{equation} 

\subsubsection{Full secular approximation (FSA):} \label{FSA}
 As a first approach we make the full secular approximation; i.e., we neglect fast rotating terms in \eref{MasterWithAnsatz} and keep only contributions where $\omega_{nm} - \omega_{kl}$ vanishes.
In this way the off-diagonal elements of $\sigma_{nm}$ are decoupled from the diagonal ones so that
\begin{eqnarray}
   \dot \sigma_{nn} (t) = \pi \sum_{k} \mathcal{L}_{nn,kk}  \sigma_{kk}(t),  \label{DiagonalFSA}\\
   \dot \sigma_{nm} (t) = \pi  \mathcal{L}_{nm,nm}  \sigma_{nm}(t) \textrm{\quad for \quad} n \neq m \label{OffDiagonalFSA}.
\end{eqnarray}
The equation for the off-diagonal elements is then
\begin{equation} \label{SolOffDiagFSA}
  \sigma_{nm}(t) = \sigma_{nm}^0 e^{\pi \mathcal{L}_{nm,nm}t},
\end{equation} 
which becomes with \eref{AnalyAnsatz}
\begin{equation}\label{SolOffDiagFSA2}
  \rho_{nm}(t) = \rho_{nm}^0 e^{\pi \mathcal{L}_{nm,nm}t} e^{-\rmi \omega_{nm} t}.
\end{equation} 
As through the FSA the oscillatory motion of the dynamics is separated from the relaxation one we can divide \eref{PtQubitHO} into two parts,
\begin{equation} \label{FSAPt}
  P(t) = P_{\rm relax.}(t) + P_{\rm dephas.}(t),
\end{equation} 
where $P_{\rm relax.} (t) = \sum_n p_{nn}(t)$ describes the relaxation and $P_{\rm dephas.}(t) = \sum_{n > m} p_{nm}(t)$ the dephasing parts of the dynamics. With \eref{SolOffDiagFSA2} the latter takes the form
\begin{equation} \label{FSADephasingPart}
  P_{\rm dephas.}(t) = \sum_{n>m} p_{nm}(0) \rme^{- \Gamma_{nm} t} \cos (\omega_{nm} t)
\end{equation} 
with the dephasing rates $\Gamma_{nm} \equiv - \pi \mathcal{L}_{nm,nm}$.
Expressions for the dephasing coefficients $\mathcal{L}_{nm,nm}$ can be found in \ref{RateCoeff} and the initial conditions $\rho_{nm}^0=\sigma_{nm}^{0} = \rho_{nm}(0)$ are given by \eref{StartingConditions}.
The diagonal elements are more difficult to obtain, as one has to solve a system of coupled differential equations,  \eref{DiagonalFSA}. 
Calculating the corresponding rate coefficients of this system for the five lowest eigenstates, we find that there are only eight independent ones, namely $\mathcal{L}_{00,11}$, $\mathcal{L}_{00,22}$, $\mathcal{L}_{11,22}$, $\mathcal{L}_{11,33}$, $\mathcal{L}_{11,44}$, $\mathcal{L}_{22,33}$, $\mathcal{L}_{22,44}$ and $\mathcal{L}_{33,44}$. They are given by
\begin{equation} \label{IndependentRateCoeff}
  \mathcal{L}_{jj,kk} = 2 G(\omega_{jk}) N_{jk} X_{jk}^2 \quad {\rm with} \quad j<k,
\end{equation} 
where $j$ and $k$ adopt the above values. Furthermore, $\mathcal{L}_{00,33}$, $\mathcal{L}_{00,44}$, $\mathcal{L}_{33,00}$ and  $\mathcal{L}_{44,00}$ vanish. The remaining rate coefficients are combinations of the above. We find that
\begin{equation} \label{DependentRateCoeff}
  \mathcal{L}_{kk,jj} = \mathcal{L}_{jj,kk} + 2 G(\omega_{jk}) X_{jk}^2 = (N_{jk}+1) 2 G(\omega_{jk}) X_{jk}^2
\end{equation} 
and
\numparts
\begin{eqnarray} 
    \mathcal{L}_{00,00} = - \mathcal{L}_{11,00} -  \mathcal{L}_{22,00}, \label{DependentRates1}\\
    \mathcal{L}_{11,11} = - \mathcal{L}_{00,11} -  \mathcal{L}_{22,11} - \mathcal{L}_{33,11} - \mathcal{L}_{44,11}, \\
    \mathcal{L}_{22,22} = - \mathcal{L}_{00,22} -  \mathcal{L}_{11,22} - \mathcal{L}_{33,22} - \mathcal{L}_{44,22}, \label{DependentRates2}\\
    \mathcal{L}_{33,33} = - \mathcal{L}_{11,33} -  \mathcal{L}_{22,33} - \mathcal{L}_{44,33}, \\
    \mathcal{L}_{44,44} = - \mathcal{L}_{11,44} -  \mathcal{L}_{22,44} - \mathcal{L}_{33,44}. \label{DependentRates3}
\end{eqnarray}
\endnumparts

However, the system \eref{DiagonalFSA} is still too complicated to be solved analytically.
Thus, we invoke a further approximation: we consider the factor
 $N_{nm}+1 = \frac{1}{2} [ \coth ( \hbar \beta \omega_{nm}/2) +1]$ with $n<m$
in \eref{DependentRateCoeff} and use that
  $\lim_{\omega \to -\infty} \coth ( \hbar \beta \omega_{nm}/2 ) = -1$.
It depends strongly on the temperature $\beta$ for which value of $\omega_{nm}$ this limit is reached approximately. For the parameters we are working with one usually is in the region where $(N_{nm}+1) \ll 1$. Thus, we will neglect in the following terms containing the factor $(N_{nm}+1)$.
Furthermore, one sees from \eref{DistanceQuasiEn} that $\omega_{12} \backsim g$ and $\omega_{34} \backsim g$. 
With that $\mathcal{L}_{11,22} = \Or(g^3)$ and $\mathcal{L}_{33,44} = \Or(g^3)$
can be neglected.
Using \eref{DependentRates1} -- \eref{DependentRates3}  the matrix of the system (\ref{DiagonalFSA}) becomes
\begin{equation} \label{SimplifiedMasterEq}
\fl \mathcal{L}_{\rm relax.} =
   \left( \begin{array}{ccccc}
                  0 & \mathcal{L}_{00,11} & \mathcal{L}_{00,22} & 0 & 0 \\
                  0 & -\mathcal{L}_{00,11} & 0 & \mathcal{L}_{11,33} & \mathcal{L}_{11,44} \\
                  0 & 0 & -\mathcal{L}_{00,22} & \mathcal{L}_{22,33} & \mathcal{L}_{22,44}\\
                  0 & 0 & 0 & - \mathcal{L}_{11,33} - \mathcal{L}_{22,33} & 0  \\
                  0 & 0 & 0 & 0 & - \mathcal{L}_{11,44} - \mathcal{L}_{22,44}
                \end{array}
       \right).
\end{equation} 
The eigenvalues and eigenvectors of this matrix and the associated time evolution of the elements $\sigma_{nn}(t)$ are given in  \eref{SolDiagFSA00} -- \eref{SolDiagFSA44} of  \ref{AppDiagonalElements}.
Unlike for the dephasing part  \eref{FSADephasingPart}, we cannot extract a simple analytical expression for the relaxation rate as $P_{\rm relax.}(t) = \sum_{n } p_{nn}(t)$ now consists of a sum of several exponential functions, cf. \eref{p0} together with \eref{SolDiagFSA00} -- \eref{SolDiagFSA44}. But still we are able to provide an analytical formula for $P(t)$ using \eref{FSAPt}.

\subsubsection{An ansatz for the long-time dynamics:} \label{AnsatzFSA}
\begin{figure}[h]
\begin{flushright}
\resizebox{.85\linewidth}{!}{\includegraphics{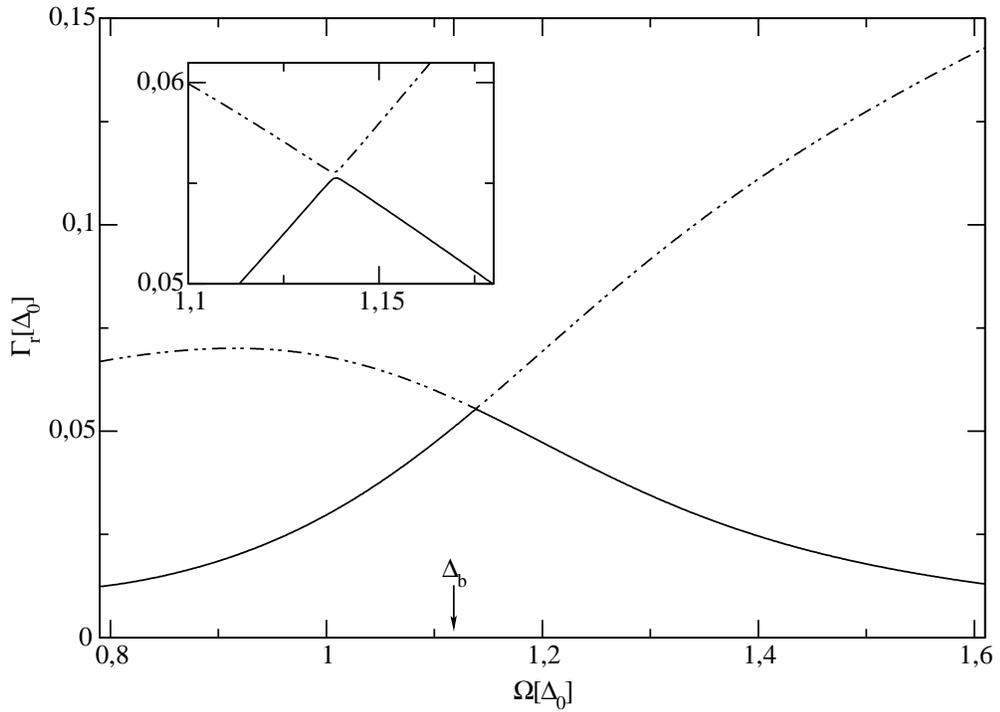}}
\end{flushright}
\caption{The relaxtion rate $\Gamma_{\rm r}$ as it is given in \eref{RelaxationRate} drawn against the oscillator frequency $\Omega$ (solid line). Used values are $\varepsilon = 0.5 \Delta_0$, corresponding to a frequency splitting $\Delta_{\rm b}= 1.118 \Delta_0$, and coupling $g=0.18 \Delta_0$. Moreover, the damping constant is $\kappa=0.0154$ and $\beta =10 (\hbar \Delta_0)^{-1}$. At resonance ($\Omega = \Delta_{\rm b}$) $\Gamma_{\rm r}$ is maximal. For a comparison also the second smallest eigenvalue is plotted (dashed line). The inset shows the two eigenvalues close to resonance. \label{RelaxationRateFig}}
\end{figure}

In order to obtain a simple expression for the relaxation part, we consider the long-time dynamics of the system. In other words, rather than looking at the many relaxation contributions to the populations $\sigma_{nn}(t)$, we focus on the smallest eigenvalue of the relaxation coefficients, as it will dominate at long times. Further, we consider only the rate matrix associated to the three lowest qubit-HO eigenstates, $\ket{0}$, $\ket{1}$ and $\ket{2}$ in \eref{DiagonalFSA} and obtain with \eref{DependentRates1} - \eref{DependentRates2}  that
\begin{equation} \label{RelaxationMatrix}
\fl  \mathcal{L}_{\rm relax.} =
  \left( \begin{array}{ccc}
                  - \mathcal{L}_{11,00} - \mathcal{L}_{22,00} & \mathcal{L}_{00,11} & \mathcal{L}_{00,22}  \\
                  \mathcal{L}_{11,00} & -\mathcal{L}_{00,11} - \mathcal{L}_{22,11} &  \mathcal{L}_{11,22} \\
                  \mathcal{L}_{22,00} & \mathcal{L}_{22,11} & -\mathcal{L}_{00,22} -\mathcal{L}_{11,22}
                \end{array}
       \right).
\end{equation} 
Here, we have not neglected the rate coefficients containing the term $(N_{nm}+1)$ and further took $\mathcal{L}_{11,22}$ into account despite of being of third order in $g$ as such contribution removes the degeneracy between the two lowest eigenvalues at resonance, cf. inset in \fref{RelaxationRateFig}. The smallest eigenvalue  reads
\begin{eqnarray} \label{RelaxationRate}
    \fl  \Gamma_{\rm r} \equiv - \frac{\pi}{2} \biggl\{ - \sum_{n\neq m} \mathcal{L}_{nn,mm}  + \biggl[  \biggl( \sum_{n\neq m} \mathcal{L}_{nn,mm}  \biggr)^2 -4 (\mathcal{L}_{00,11} \mathcal{L}_{00,22} + \mathcal{L}_{11,00} \mathcal{L}_{00,22}   \nonumber \\
     + \mathcal{L}_{00,11} \mathcal{L}_{11,22} + \mathcal{L}_{11,00} \mathcal{L}_{11,22} + \mathcal{L}_{00,11} \mathcal{L}_{22,00} + \mathcal{L}_{11,22} \mathcal{L}_{22,00} \nonumber\\
     + \mathcal{L}_{22,11} \mathcal{L}_{00,22} + \mathcal{L}_{11,00} \mathcal{L}_{22,11} + \mathcal{L}_{22,00} \mathcal{L}_{22,11} ) \biggr]^{1/2} \biggr\}.
\end{eqnarray}
With the system being detuned this expression can be simplified further, namely
  \begin{equation} \label{ApproximativeRelax}
     \Gamma_{\rm r} \approx \pi \mathcal{L}_{00,22} \quad {\rm for} \quad \Omega < \Delta_{\rm b}; \quad 
     \Gamma_{\rm r} \approx \pi \mathcal{L}_{00,11} \quad {\rm for} \quad \Omega > \Delta_{\rm b} \nonumber.
  \end{equation}  
In \fref{RelaxationRateFig} the relaxation rate $\Gamma_{\rm r}$ as obtained from \eref{RelaxationRate} is shown as a function of the oscillator frequency $\Omega$. Clearly, it is maximal at resonance ($\Omega = \Delta_{\rm b}$), whereas it decays with $\Omega$ being detuned from the resonance. This effect has already been predicted by Blais \etal \cite{Blais}. As the qubit is not directly coupled to the bath but rather through the oscillator, the latter being detuned filters out the environmental noise at the qubit transition frequency. Additionally, we show the second smallest eigenvalue of \eref{RelaxationMatrix}. We notice that close to the resonant point ($\Omega = \Delta_{\rm b}$) there is an avoided crossing.
Finally, we find that
\begin{equation}
  P_{\rm relax.}(t) = (p_0 - p_\infty) e^{-\Gamma_{\rm r}t} + p_\infty, 
\end{equation} 
where like in \sref{DynamicsNoDiss} $p_0 \equiv \sum_n p_{nn}(0)$. For getting $p_\infty$ we have in principle to find the steady-state solution of \eref{DiagonalFSA}. Here, we just assume for $t \to \infty$ a Boltzmann distribution for the qubit-HO system, so that
  $\rho_{nn}(\infty) = Z_{\rm QHO}^{-1} \rme^{-\beta E_n}$ with $Z_{\rm QHO} = \sum_{n} \rme^{-\beta E_{n}}$.
Thus,
\begin{equation}
  \fl p_\infty = \sum_n \sum_i \left\{ \cos \Theta \biggl[ \braket{j {\rm g}}{n}^2 - \braket{j {\rm e}}{n}^2 \biggr]  + 2 \sin \Theta \braket{j {\rm g}}{n} \braket{j {\rm e}}{n} \right\} \rho_{n n}(\infty) \label{pinf}.
\end{equation}
The formula for the long-time dynamics is obtained as
        \begin{equation} \label{PSimpleFSA}
          P(t) = (p_0 - p_\infty) e^{-\Gamma_{\rm r}t} + p_\infty + \sum_{n>m} p_{nm}(0) e^{-\Gamma_{nm} t} \cos(\omega_{nm} t).
        \end{equation}
To get further insight on the dominant frequencies we evaluate the Fourier transform of \eref{PSimpleFSA} according to
\begin{equation}
  F(\omega) = 2 \int_0^\infty dt \cos \omega t P(t),
\end{equation} 
yielding
\begin{eqnarray} \label{FSimpleFSA}
 \fl  F(\omega) = 2 (p_0 - p_\infty) \frac{\Gamma_{\rm r}}{\omega^2 + \Gamma_{\rm r}^2} + 2 \pi p_\infty \delta(\omega) \nonumber\\
  + \sum_{n<m} p_{nm} \Gamma_{mn} \left[ \frac{1}{\Gamma_{mn}^2 + (\omega_{mn} + \omega)^2} + \frac{1}{\Gamma_{mn}^2 + (\omega_{mn} - \omega)^2}   \right].
\end{eqnarray}

\subsubsection{Partial secular approximation (PSA):} \label{PSA}
An improvement to the FSA is to take into account certain non-vanishing contributions of $\omega_{nm} - \omega_{kl}$.  We have to keep in mind, that  there are quasi-degenerate levels close to resonance. In our case the first with second energy level and the third with fourth one build a doublet, meaning that they are close together in energy space. The level spacing is approximately proportional to $g$ for the former and $\sqrt{2} g$ for the latter. Because of that and as the transitions from level three and four are less probable, we will in the following only consider the first and second level as being almost degenerate. %As a consequence the difference between the frequencies $\omega_{10}$ and $\omega_{20}$, $\omega_{31}$ and $\omega_{32}$ or $\omega_{41}$ and $\omega_{42}$ are  relatively small. Furthermore $\omega_{21}$ is small.
Taking this into account  in  \eref{MasterWithAnsatz} we arrive for the diagonal elements at
\begin{equation}
  \fl \dot \sigma_{nn}(t) = \pi \sum_k \mathcal{L}_{nn,kk}\sigma_{kk}(t) + \pi \mathcal{L}_{nn,12} \sigma_{12} (t) \rme^{-\rmi\omega_{12} t}+ \pi \mathcal{L}_{nn,21} \sigma_{21} (t) \rme^{-\rmi\omega_{21} t}.
\end{equation} 
A numerical analysis shows that the effect of the last two terms on the right-hand side of the above equation will in the worst case lead to very small wiggles in $\sigma_{nn}(t)$ and play no role in $P(t)$. Thus, we finally write
\begin{equation}\label{DiagonalPSA}
   \dot \sigma_{nn}(t)\cong  \pi \sum_k \mathcal{L}_{nn,kk}\sigma_{kk}(t),
\end{equation} 
which is the same equation as we got in the FSA approach. %That means that for the diagonal elements we can use the above calculations.
 However, the off-diagonal contributions $\sigma_{01}$, $\sigma_{02}$, $\sigma_{13}$, $\sigma_{23}$, $\sigma_{14}$ and $\sigma_{24}$ have to be examined more carefully. From  (\ref{MasterWithAnsatz}) we find that one has to solve the equations
\begin{eqnarray}
  \dot \rho_{nm} (t) = (- \rmi \omega_{nm} + \pi \mathcal{L}_{nm,nm}) \rho_{nm}(t) + \pi \mathcal{L}_{nm,jk} \rho_{jk}(t), \\
  \dot \rho_{jk} (t) = \pi \mathcal{L}_{jk,nm} \rho_{nm}(t) + (\rmi \omega_{jk} + \pi \mathcal{L}_{jk,jk}) \rho_{jk}(t)
\end{eqnarray}
with $\{ (nm), (jk) \} = \{(01);(02)\}$,$\{(13);(23)\}$ or $\{(14);(24)\}. $
As solution one gets
\begin{eqnarray} 
  \rho_{nm} = c_{nm,jk}^{(+)} v_{nm,jk}^{(+)} \rme^{\lambda_{nm,jk}^{(+)} t} + c_{nm,jk}^{(-)} v_{nm,jk}^{(-)} \rme^{\lambda_{nm,jk}^{(-)} t}, \label{OffDiaPSA1}\\
  \rho_{jk} = c_{nm,jk}^{(+)} \rme^{\lambda_{nm,jk}^{(+)} t} + c_{nm,jk}^{(-)} \rme^{\lambda_{nm,jk}^{(-)} t}. \label{OffDiaPSA2}
\end{eqnarray}
Here, the oscillation frequencies and the decay of the off-diagonal elements are given by
\begin{equation} \label{DephasingEigenvalues}
  \lambda_{nm,jk}^{(+/-)} = \frac{1}{2} \left[ \pi (\mathcal{L}_{nm,nm} + \mathcal{L}_{jk,jk}) - \rmi (\omega_{nm} + \omega_{jk}) \pm R_{nm,jk}   \right]
\end{equation} 
with
\begin{equation}
  \fl R_{nm,jk} = \sqrt{\left[ \pi (\mathcal{L}_{nm,nm} - \mathcal{L}_{jk,jk}) - \rmi (\omega_{nm} - \omega_{jk}) \right]^2 + 4 \pi^2 \mathcal{L}_{nm,jk} \mathcal{L}_{jk,nm}}.
\end{equation} 
The amplitudes of the oscillations are given through the coefficients
\begin{equation}
  \fl c_{nm,jk}^{(+/-)} = \pm \frac{2 \pi \mathcal{L}_{jk,nm} \rho_{nm}^0 - \rho_{jk}^0 \left[ \pi (\mathcal{L}_{nm,nm} - \mathcal{L}_{jk,jk}) - \rmi (\omega_{nm} - \omega_{jk}) \mp R_{nm,jk} \right]}{2 R_{nm,jk}}
\end{equation}
and
\begin{equation}
  \fl v_{nm,jk}^{(+\-)} = \frac{2 \pi}{\mathcal{L}_{jk,nm}} \left[ \pi (\mathcal{L}_{nm,nm} - \mathcal{L}_{jk,jk}) - \rmi (\omega_{nm} - \omega_{jk}) \pm R_{nm,jk} \right].
\end{equation}
\begin{figure}[h]
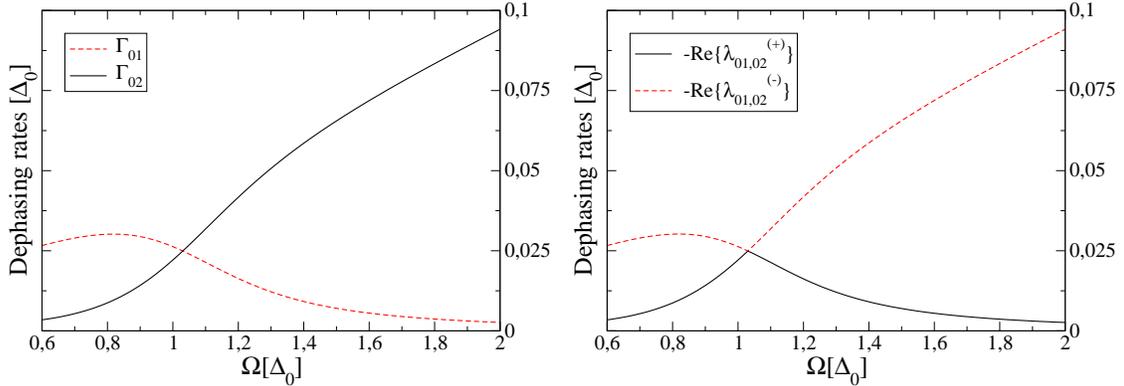

\begin{flushright}
\resizebox{.47\linewidth}{!}{
  \includegraphics{Figures/figure4a.eps}}
\resizebox{.47\linewidth}{!}{
  \includegraphics{Figures/figure4b.eps}}
\caption{Comparison between the dephasing rates of the two dominant frequencies as they are obtained using the FSA or the PSA, respectively. On the left: the FSA rates $\Gamma_{01} \equiv - \pi \mathcal{L}_{01,01}$ (dashed red line) and $\Gamma_{02} \equiv - \pi \mathcal{L}_{02,02}$ (solid black line). On the right: the real part of $\lambda_{01,02}^{(-)}$ (red dashed line) and $\lambda_{01,02}^{(+)}$ (black solid line)  as given by \eref{DephasingEigenvalues} is shown. The rate dominating the dephasing behaviour is defined as $\Gamma_{12}^{(+)} \equiv \Re\{\lambda_{01,02}^{(+)}\}$.  For $\Omega < \Delta_0$ we see that $\Gamma_{12}^{(+)}$ is approximated by the FSA rate $ \Gamma_{02}$, while for $\Omega > \Delta_0$ by $\Gamma_{01}$. Used values are $\varepsilon = 0$, $g=0.18$, $\kappa = 0.0154$ and $\beta = 10 (\hbar \Delta_0)^{-1}$.  \label{FigDephasingRates}}
\end{flushright}
\end{figure}
Thus, we have again all ingredients to calculate analytically the relaxation and dephasing part of \eref{FSAPt}. For the PSA we cannot provide a simple expression for the dephasing rates as in the FSA, where we had $\Gamma_{nm} = -\pi \mathcal{L}_{nm,nm}$. As one can see from \eref{OffDiaPSA1} and \eref{OffDiaPSA2}, $\rho_{01}$ and $\rho_{02}$ are a mixture of  contributions decaying with $\Re\{\lambda_{01,02}^{(+)}\}$ and $\Re\{\lambda_{01,02}^{(-)}\}$. Similar to our findings for the relaxation rate, also here the smallest eigenvalue will dominate the dephasing behaviour. From the right graph in \fref{FigDephasingRates} we find that this is $\Re\{ \lambda_{01,02}^{(+)}\} \equiv \Gamma_{12}^{(+)}$. Comparing it with the dephasing rates we got using the FSA, left graph in \fref{FigDephasingRates}, we see that for negative detuning ($\Omega < \Delta_{\rm b}$) the rate $\Gamma_{02} = - \pi \mathcal{L}_{02,02}$ approximates $\Gamma_{12}^{(+)}$, whereas for positive detuning ($\Omega > \Delta_{\rm b}$) this is done by $\Gamma_{01} = - \pi \mathcal{L}_{01,01}$. In the FSA $\Gamma_{02}$ and $\Gamma_{01}$ correspond to the frequencies $\omega_{10}$ and $\omega_{20}$, respectively. In the PSA the frequency $\omega_{12}^{(+)} \equiv \Im\{ \lambda_{01,02}^{(+)} \}$ is given by $\omega_{12}^{(+)} = \omega_{20}$ for $\Omega < \Delta_{\rm b}$ and $\omega_{12}^{(+)} = \omega_{10}$ for $\Omega > \Delta_{\rm b}$. Hence, for negative detuning oscillations with frequency $\omega_{20}$ will dominate the dynamics, while those with $\omega_{10}$ will almost vanish. For positive detuning it is the other way round. In \eref{OffDiaPSA1} and \eref{OffDiaPSA2} this behaviour is reflected by the coefficients $c_{nm,jk}^{(+/-)}$ and $v_{nm,jk}^{(+/-)}$. Around resonance ($\Omega \approx \Delta_{\rm b}$) the PSA tells us by \eref{OffDiaPSA1} and \eref{OffDiaPSA2} that the dephasing rates and frequencies are a mixture of $\Gamma_{01}$ and $\Gamma_{02}$ or $\omega_{10}$ and $\omega_{20}$, respectively. From the left graph in \fref{FigDephasingRates} one notices further that the FSA rate $\Gamma_{02}$ grows linearly with $\Omega$ for positive detuning. However, as the weight of the corresponding frequency $\omega_{20}$ will be almost zero, $\Gamma_{02}$ will give no relevant contribution to $P_{dephas.}(t)$ in this regime but the dephasing  will rather be associated to the FSA rate $\Gamma_{01}$. Hence, out of resonance the FSA will still fairly well describe the dynamics of $P(t)$. Comparing the expressions for $\mathcal{L}_{01,01}$ and $\mathcal{L}_{02,02}$ given in \ref{RateCoeff} by \eref{L0101} and \eref{L0202} with the approximative expressions for the relaxation rate at positive and negative detuning \eref{ApproximativeRelax}, we see that for zero bias ($\varepsilon =0 $) the PSA dephasing rate is equal to $\Gamma_{\rm r}/2$. For a biased system an additional term is added depending on the spectral density of the bath at $\omega =0$.
\begin{figure}[h]
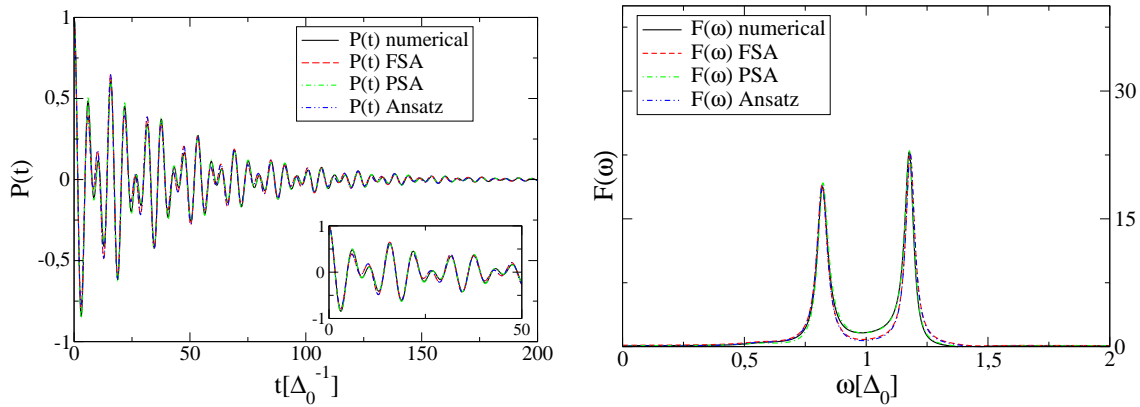

\begin{flushright}
\resizebox{0.47\linewidth}{!}{
 \includegraphics{Figures/figure5a.eps}}
\hspace{0.1 cm}
\resizebox{0.47\linewidth}{!}{
 \includegraphics{Figures/figure5b.eps}}
\end{flushright}
\caption{Comparison between the behaviour of $P(t)$ and its Fourier transform $F(\omega)$ as obtained from the numerically exact solution (black solid curve) of the equation \eref{MasterEqQOBasis}  and the three analytical approximations discussed in the text.  The red dashed curve is the full secular approximation (FSA) solution, the green dotted-dashed curve the partial secular approximation (PSA) solution and the blue double-dotted-dashed curve the analytical formulas \eref{PSimpleFSA} and \eref{FSimpleFSA}. The parameters are $\varepsilon = 0$, $\Omega =\Delta_0$, $g=0.18 \Delta_0$, $\kappa = 0.0154$ and $\beta =10 (\hbar \Delta_0)^{-1}$. For the choosen regime of parameters differences between numerical and analytical results are barely visible. \label{Comparison_PTFW}}
\end{figure}
In \fref{Comparison_PTFW} we compare the three analytical solutions described above to the numerical solution of the master equation for the case of an unbiased TLS being at resonance with the oscillator. Concerning both the dynamics of $P(t)$ and its Fourier spectrum we see a good agreement between the different solutions. The one being closest to the numerical solution is the PSA solution. We also want to mention that going to stronger damping $\kappa$, the FSA results start to show deviations from the numerical solution. Here, one should use the PSA only. However, for the parameter regime used in the following, we will mainly apply \eref{PSimpleFSA} due to its simple, analytical form.

\section{Discussion of the results} \label{DiscussionQubits}
Having solved the master equation \eref{MasterEqQOBasis} analytically and numerically we can examine the dynamics of the system and its Fourier transform for different situations. First, we will look at a qubit operated at the degeneracy point ($\varepsilon = 0$) being in and out of resonance with the oscillator. Then, we will concentrate on the biased qubit in the same regime of parameters.

\subsection{The unbiased qubit} \label{UnbiasedQubit}
\begin{figure}[h]
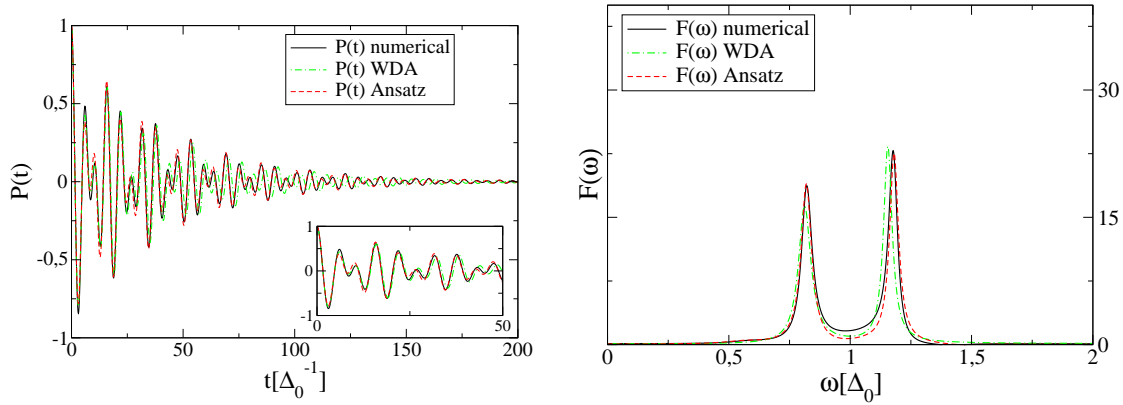

\begin{flushright}
\resizebox{0.47\linewidth}{!}{
 \includegraphics{Figures/figure6a.eps}}
\hspace{0.1 cm}
\resizebox{0.47\linewidth}{!}{
 \includegraphics{Figures/figure6b.eps}}
\end{flushright}
\caption{Dynamics of the population difference $P(t)$ (left-hand side) and its Fourier transform $F(\omega)$ (right-hand side) for an unbiased TLS being in resonance with the oscillator ($\Omega = \Delta_0$). The numerical solution of the master equation (black solid line) is compared with the analytical expressions \eref{PSimpleFSA} and \eref{FSimpleFSA} (red dashed line) and the weak damping approximation (WDA) from \cite{Nesi} (green dotted-dashed line). The parameters are $\varepsilon = 0$, $\Omega = \Delta_0$, $g=0.18 \Delta_0$, $\kappa = 0.0154$ and $\beta = 10 (\hbar \Delta_0)^{-1}$. From the Fourier transform one sees that two frequencies are dominating the dynamics. The separation of those two peaks is approximately $2 g$. The non-dissipative dynamics is shown for comparison in \fref{PtNoBiasRes}. \label{FigUnbiasedRes}}
\end{figure}

\begin{figure}[h]
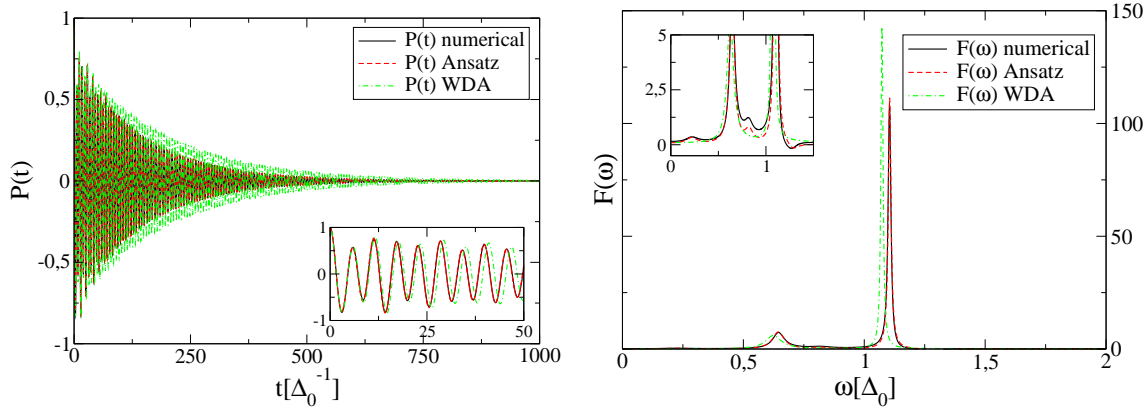

\begin{flushright}
\resizebox{0.47\linewidth}{!}{
 \includegraphics{Figures/figure7a.eps}}
\hspace{0.1 cm}
\resizebox{0.47\linewidth}{!}{
 \includegraphics{Figures/figure7b.eps}}
\end{flushright}
\caption{Dynamics of $P(t)$ and its Fourier transform $F(\omega)$ for negative detuning ($\Omega < \Delta_0$) and for $\varepsilon = 0$. Same parameters as in \fref{FigUnbiasedRes} are used except that now $\Omega = 0.75 \Delta_0$. The frequency $\omega_{20}$ dominates the dynamics. The inset on the right graph shows a zoom into the Fourier transform. The  numerical solution and the analytical expression \eref{FSimpleFSA} exhibit besides the main peaks at $\omega_{10}$ and $\omega_{20}$ two additional peaks, corresponding to the frequencies $\omega_{24}$  (between the two main peaks) and $\omega_{23}$ (on the left of the first main peak). The two dips come from $\omega_{13}$ and $\omega_{14}$.  \label{FigUnbiasedNegDetun}}
\end{figure}

For unbiased qubits we can compare our predictions with the analytical results obtained in \cite{Nesi} by starting from  a spin-boson model with the effective spectral density \eref{EffDens}. In \cite{Nesi} a so-called weak damping approximation (WDA) based on the non-interacting blip approximation (NIBA) is applied. The WDA allows a non-perturbative treatment of the coupling between the TLS and HO and hence can reproduce the occurence of two dominating frequencies as expected e.g. from exact QUAPI calculations \cite{Thorwart}. The NIBA, and hence the WDA, however, become not reliable for a biased TLS. We find that the overall agreement between our approach and the WDA is very good.  However, in the WDA solution the frequencies are slightly shifted compared to the ones obtained from our master equation. This may result from the perturbative expansion we have performed with respect to $g$ by applying the Van-Vleck perturbation theory.
\newline
First,  we look at the resonant case shown in \fref{FigUnbiasedRes}.
In agreement with previous works \cite{Thorwart, Nesi}, we find that the dynamics is dominated by two frequencies corresponding to $\omega_{10}$ and $\omega_{20}$ with separation being approximately $2g$. The weight of the latter is a bit larger.  
The reason for the bigger weight is that at resonance ($\Omega = \Delta_{\rm b}$) the qubit-HO eigenstate $\ket{j}$ is not a symmetric superposition of the states $\ket{j,\rme}$ and $\ket{j+1,{\rm g}}$ unlike it is predicted by the  Janyes-Cummings model (cf e.g. \cite{Blais}). We notice that the two unequal peaks have indeed been experimentally observed in \cite{Wallraff} (see Fig. 4b therein). Considering the states $\ket{1}_{\rm eff}$ and $\ket{2}_{\rm eff}$ in \eref{VanVleckEigenstate1} and \eref{VanVleckEigenstate2}, one already sees that for a symmetric superposition of these states we need that $\delta_0$ vanishes or that $\Omega \equiv [(\Delta_{\rm b}^4 + 2 g^2 \Delta_0^2)/\Delta_{\rm b}^2]^{-\frac{1}{2}}$ (cf . \eref {BlochSiegert}). Besides, in order to get the qubit-HO eigenstates one still has to perform the Van-Vleck transformation, which adds contributions to $\ket{1}$ and $\ket{2}$ from states corresponding to oscillator levels higher than $j=1$. Thus, our system behaves for $\Omega = \Delta_0$ as being negatively detuned, which means that the peak belonging to the higher frequency dominates, as we will show below. Slightly increasing $\Omega$ will give a stronger weight to the peak at $\omega_{10}$. This effect is not very pronounced for the non-dissipative dynamics of the unbiased qubit (\fref{PtNoBiasRes}), as there the two frequencies are still almost equally weighted. Looking however at the Fourier transform of the dissipative dynamics \eref{FSimpleFSA}, one notices that the relaxation rate also contributes to the weight of the peaks with a prefactor $\Gamma_{nm}^{-1}$. As for a negative detuned system $\Gamma_{01}$ is slightly bigger than $\Gamma_{02}$, the difference between the two peaks becomes more clear in the dissipative case. For $\varepsilon \neq 0$ the effect can already be noticed in the non-dissipative case.\newline
Next, we consider in \fref{FigUnbiasedNegDetun} the case of negative detuning, where $\Omega < \Delta_0$. No matter which approach one is looking at, clearly the frequency $\omega_{20}$ is dominating. Furthermore, paying attention to the timescale of the dynamics, one notices that the relaxation time is {\it enhanced} compared to the one we found for the resonant system. This behaviour was already explained by the formula \eref{RelaxationRate} for the relaxation rate. Again, the numerical and the solution obtained by using the long-time ansatz in \sref{AnsatzFSA} agree quite well with each other, whereas the amplitude of the oscillation with frequency $\omega_{20}$ is stronger in the WDA approach. 
Also remarkable is the fact that looking at the Fourier transform in \fref{FigUnbiasedNegDetun} one sees in the inset already small contributions of the higher oscillator levels. The transitions corresponding to $\omega_{24}$ and $\omega_{23}$ give raise to small additional peaks, while the contributions of $\omega_{13}$ and $\omega_{14}$ are negatively weighted and cause dips. The WDA approach does not show this additional contributions. They are, however, confirmed by the numerical QUAPI calculations in \cite{Thorwart} (see figure 2 therein).
\begin{figure}[h]
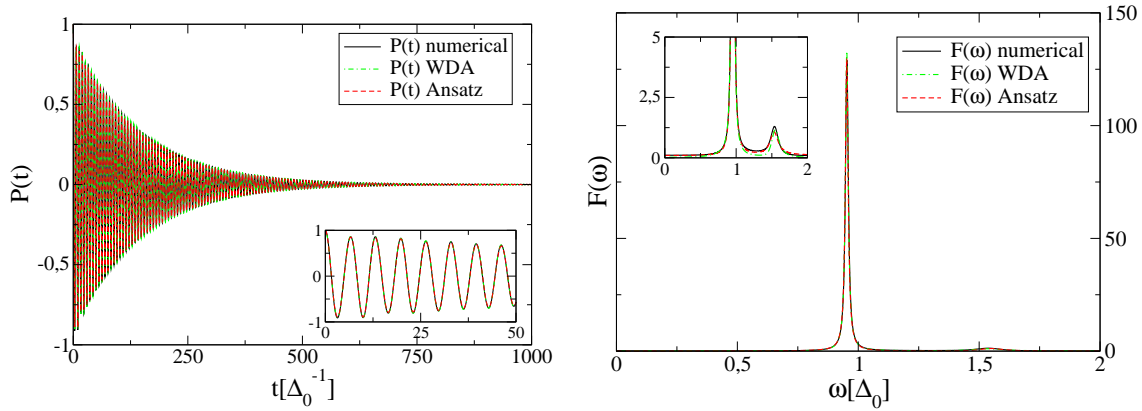

\begin{flushright}
\resizebox{0.47\linewidth}{!}{
 \includegraphics{Figures/figure8a.eps}}
\hspace{0.1 cm}
\resizebox{0.47\linewidth}{!}{
 \includegraphics{Figures/figure8b.eps}}
\end{flushright}
\caption{Dynamics of $P(t)$ and its Fourier transform $F(\omega)$ for positive detuning ($\Omega > \Delta_0$), for $\varepsilon =0$ and $\Omega = 1.5 \Delta_0$. The peak at $\omega_{10}$ dominates. No additional peaks are found. A very good agreement between all approaches discussed in the text is found. Remaining parameters are as in \fref{FigUnbiasedRes}. \label{FigUnbiasedPosDetun}}
\end{figure}
In the case of positive detuning ($\Omega>\Delta_0$) shown in \fref{FigUnbiasedPosDetun} we find a quite good agreement between all three approaches. Also for postive detuning the relaxation time is enhanced compared to the resonant case. Contrary to the negatively detuned situation the additional peaks have vanished. Besides, now the frequency $\omega_{10}$ is dominating the dynamics. This behaviour, namely that for negative detuning $\omega_{20}$ and for positive detuning $\omega_{10}$ is dominating, was already found in \cite{Thorwart}.\newline
 We will briefly explain how one can explain this observation physically. 
\begin{figure}[h]
\begin{flushright}
\rotatebox{-90}{\resizebox{!}{.84\linewidth}{
 \includegraphics{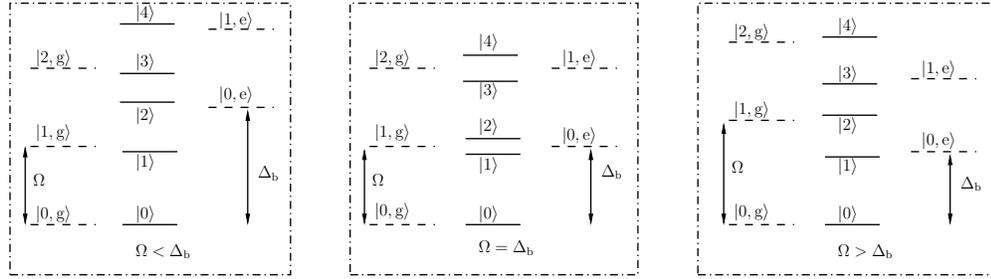}}}
\end{flushright}
\caption{Schematic energy spectrum for three different situations. From left to right: negative detuning ($\Omega < \Delta_{\rm b}$), resonant case ($\Omega = \Delta_{\rm b}$) and positive detuning ($\Omega > \Delta_{\rm b}$). The dashed lines show the energy levels for the uncoupled qubit-HO system ($g=0$). The solid lines depict the eigenstates obtained by Van-Vleck perturbation theory. \label{FigEnergySchemata}}
\end{figure}
For this we look at \fref{FigEnergySchemata}. For a detuned system ($\Omega \neq \Delta_{\rm b}$) the qubit-HO eigenstates are not symmetric superpositions of the states $ \ket{j  \rm g} $ and $ \ket{j  \rme} $. They rather asymptotically approach the eigenstates of the uncoupled qubit-HO Hamiltonian. In \fref{EnSpecCoup} we see that for a negatively detuned system (line a) the qubit-HO eigenstate $\ket{2j + 1}$ approaches the state $\ket{(j+1) \rm g}$, whereas the main contribution to the state $\ket{2j+2}$ will come from the state $\ket{j \rme}$. From the left diagram in \fref{FigEnergySchemata} we see that the state $\ket{2}$ is energetically higher than the state $\ket{1}$. However, due to the Boltzmann distributed occupation of the oscillator, the state $\ket{0 \rme}$ will be more populated than the state $\ket{1 \rm g}$ and consequently also $\ket{2}$ will exhibit a larger population than $\ket{1}$, as the latter only feels a small contribution from the state $\ket{0 \rme}$. Thus, transitions from $\ket{2}$ to the groundstate are more likely to occur than those from $\ket{1}$ to the groundstate. This explains the dominance of $\omega_{20}$ in \fref{FigUnbiasedNegDetun} and \fref{FigBiasedNegDetun}. In this case the frequency $\omega_{20} \approx \Delta_b$ and $\omega_{10} \approx \Omega$. As far as not excluded by selection rules, minor peaks from transitions to the levels lying in between can be also seen.\newline
For positive detuning (line c in \fref{EnSpecCoup}) $\ket{2j + 1}$ approaches $\ket{j \rme}$, while $\ket{2j+2}$ is close to $\ket{(j+1) \rm g}$. From the right graph in \fref{FigEnergySchemata} we see that the state $\ket{1}$, being lowest in energy apart from the groundstate, is now also more probable to be occupied than $\ket{2}$. Therefore, as confirmed by \fref{FigUnbiasedPosDetun} and \fref{FigBiasedPosDetun}, the frequency $\omega_{10}$ is dominating whereas $\omega_{20}$ is represented only by a small peak in the Fourier spectrum. Furthermore, as there are no additional energy levels between the state $\ket{1}$, which is most probably to be populated, and the ground level, other transitions than those corresponding to $\omega_{10}$ or $\omega_{20}$ are very unlikely to occur. In \fref{FigBiasedPosDetun} the dip corresponding to $\omega_{21}$ appears only very faintly.

\subsection{The biased qubit}
\begin{figure}[h]
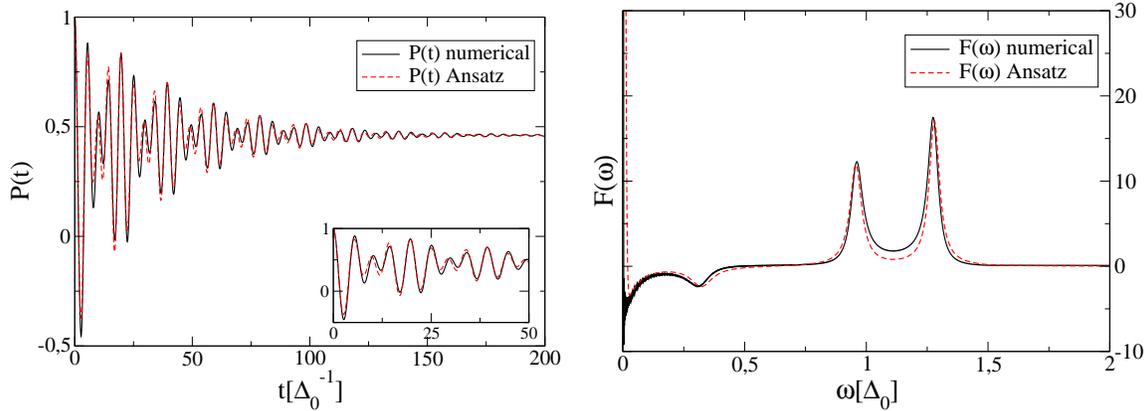

\begin{flushright}
\resizebox{0.47\linewidth}{!}{
 \includegraphics{Figures/figure10a.eps}}
\hspace{0.1 cm} 
 \resizebox{0.47\linewidth}{!}{
 \includegraphics{Figures/figure10b.eps}}
\end{flushright}
\caption{Dynamics of $P(t)$ and its Fourier transform $F(\omega)$ for the biased qubit being in resonance with the oscillator. Here, $\varepsilon = 0.5 \Delta_0$ and $\Omega = \Delta_{\rm b}$. The remaining parameters are the same as for the unbiased qubit. In the Fourier spectrum the frequenices $\omega_{10}$ and $\omega_{20}$ dominate. At frequency $\omega_{21}$ a small dip can be seen. At $\omega = 0$ the spectrum exhibits a relaxation peak. \label{FigBiasedResonant}}
\end{figure}

We will now examine a qubit being operated at finite bias. We consider the case $\varepsilon > 0$. For negative bias-offset the behaviour is analogous. Again three different situations are taken into account: the qubit being in resonance with the oscillator ($\Delta_{\rm b} = \Omega$), negative ($\Omega < \Delta_{\rm_b}$) and positive ($\Omega > \Delta_b$) detuning.\newline
For the resonant case ($\Omega = \Delta_{\rm b}$) depicted in \fref{FigBiasedResonant} we see a similiar behaviour as for the unbiased qubit. Again two frequencies, $\omega_{10}$ and $\omega_{20}$, are dominating the dynamics.
Left to the peak at $\omega_{10}$ a small dip can be found in the Fourier spectrum. This corresponds to the transition $\omega_{21}$. For infinite time the dynamics relaxes to an equilibrium value which is nonzero in contrast to the unbiased case. This can be seen in the Fourier spectrum through a relaxation peak at $\omega=0$. The peak arises because of the term
\begin{equation} \label{RelaxationPeak}
 2 (p_0 - p_\infty) \frac{\Gamma_{\rm r}}{\omega^2 + \Gamma_{\rm r}^2} + 2 \pi p_\infty \delta(\omega)
\end{equation} 
in \eref{FSimpleFSA}.  The first part of this sum gives rise to the negative shift of this peak. The reason that for the analytical solution the peak is not as strongly shifted as for the numerical one is technical: in order to plot the delta function in \eref{RelaxationPeak} we gave it a finite width, which surpresses the negative contribution of  the first term in \eref{RelaxationPeak}. Like for the unbiased qubit the highest energy level playing a role for the dynamics is $E_2$; i.e., only the ground and first excited level of the oscillator are of importance.
\begin{figure}[h]
\begin{flushright}
\resizebox{0.47\linewidth}{!}{
 \includegraphics{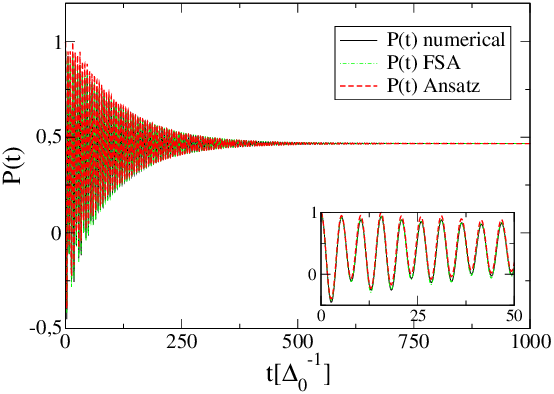}}
\hspace{0.1 cm}
\resizebox{0.47\linewidth}{!}{
 \includegraphics{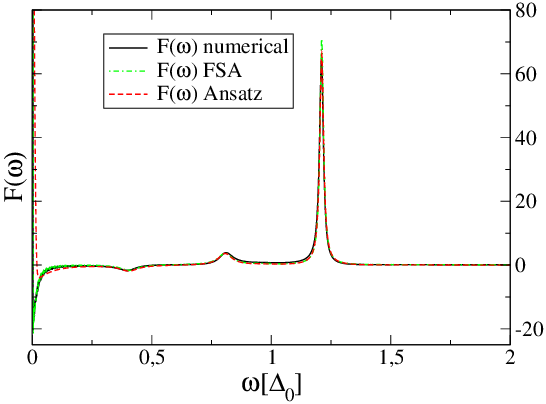}}
\end{flushright}
\caption{Dynamics of $P(t)$ and its Fourier transform $F(\omega)$ for negative detuning ($\Omega < \Delta_{\rm b}$) with $\Omega = 0.9$ and $\varepsilon = 0.5$. Next to the numerical solution (black solid curve) of the full master equation and the FSA solution \eref{PSimpleFSA} and \eref{FSimpleFSA} (red dashed curve), also the improved FSA solution of \eref{DiagonalFSA} and \eref{OffDiagonalFSA} (green dotted-dashed curve) are shown. The dynamics is dominated by $\omega_{20}$. The peak at $\omega_{10}$ is much weaker. Like for the resonant, biased qubit a dip is found at $\omega_{21}$ and a relaxation peak at $\omega =0$. \label{FigBiasedNegDetun}}
\end{figure}
In \fref{FigBiasedNegDetun} the dynamics and its Fourier transform for a negatively detuned qubit-oscillator system with $\varepsilon \neq 0$ are shown. Like for the unbiased case detuning gives raise to longer relaxation times for the qubit. Also in agreement with the unbiased case is the dominance of the frequency $\omega_{20}$. We see that for small $t$ the long-time solution \eref{PSimpleFSA} slightly overestimates the maxima of the oscillations and underestimats its minima. Furthermore, we get here the unphysical situation that the maximum of the third oscillation in $P(t)$ exceeds the value of one. The reason for that behaviour is that, by construction, we underestimate with \eref{PSimpleFSA} the relaxation at short times. As for certain paramteres the term $(p_0 - p_\infty)$ in \eref{PSimpleFSA} can become negative, it increases too fast towards the equilibrium and gives thus raise to the observed deviations in the short time behaviour. On a longer timescale both graphs agree quite well.
\begin{figure}[h]
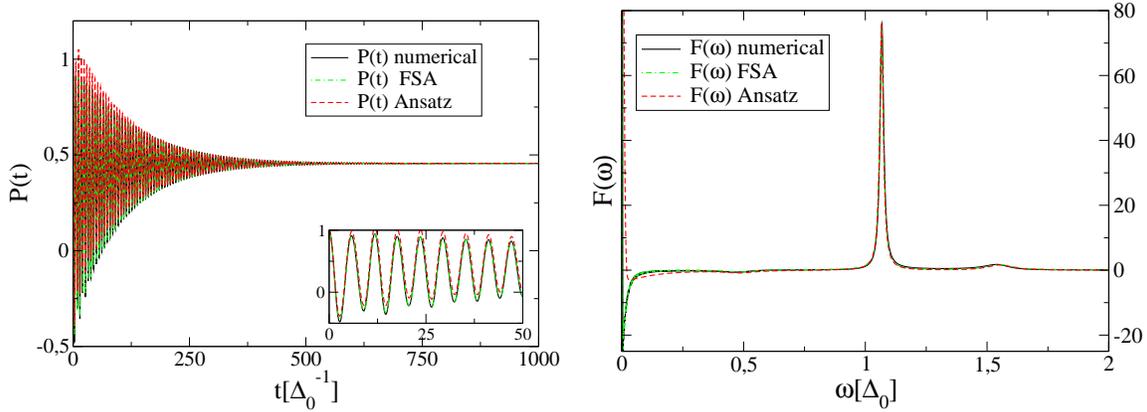

\begin{flushright}
\resizebox{0.47\linewidth}{!}{
 \includegraphics{Figures/figure12a.eps}}
\hspace{0.1 cm}
\resizebox{0.47\linewidth}{!}{
 \includegraphics{Figures/figure12b.eps}}
\end{flushright}
\caption{Dynamics and Fourier transform for positve detuning ($\Omega > \Delta_{\rm b}$) for $\Omega =1.5 \Delta_0$ and $\varepsilon = 0.5 \Delta_0$. Like in \fref{FigBiasedNegDetun} three different approaches are compared. In all three cases the frequency $\omega_{10}$ dominates. \label{FigBiasedPosDetun}}
\end{figure}
For the case of positive detuning ($\Omega > \Delta_{\rm b}$), which is presented in \fref{FigBiasedPosDetun}, the upward shift of the dynamics obtained from \eref{PSimpleFSA} and \eref{FSimpleFSA} compared to the numerical graph of $P(t)$ at small times is even stronger.  To visualize that it is not a failure of the FSA approach we show in \fref{FigBiasedNegDetun} and \fref{FigBiasedPosDetun} additionally the analytical FSA solution of \eref{DiagonalFSA} and \eref{OffDiagonalFSA} calculated in \sref{FSA}. The latter agrees very well with the numerical solution.  At long time-scales and for the Fourier spectrum all three approaches match with each other very well.\newline
To conclude this paragraph we want to mention that all the results found both for the unbiased and the biased qubit confirm the numerical QUAPI results in \cite{Thorwart}.

\subsection{Symmetrized correlation function}
\begin{figure}[h]
\begin{flushright}
\rotatebox{-90}{
\resizebox{!}{0.47\linewidth}{
 \includegraphics{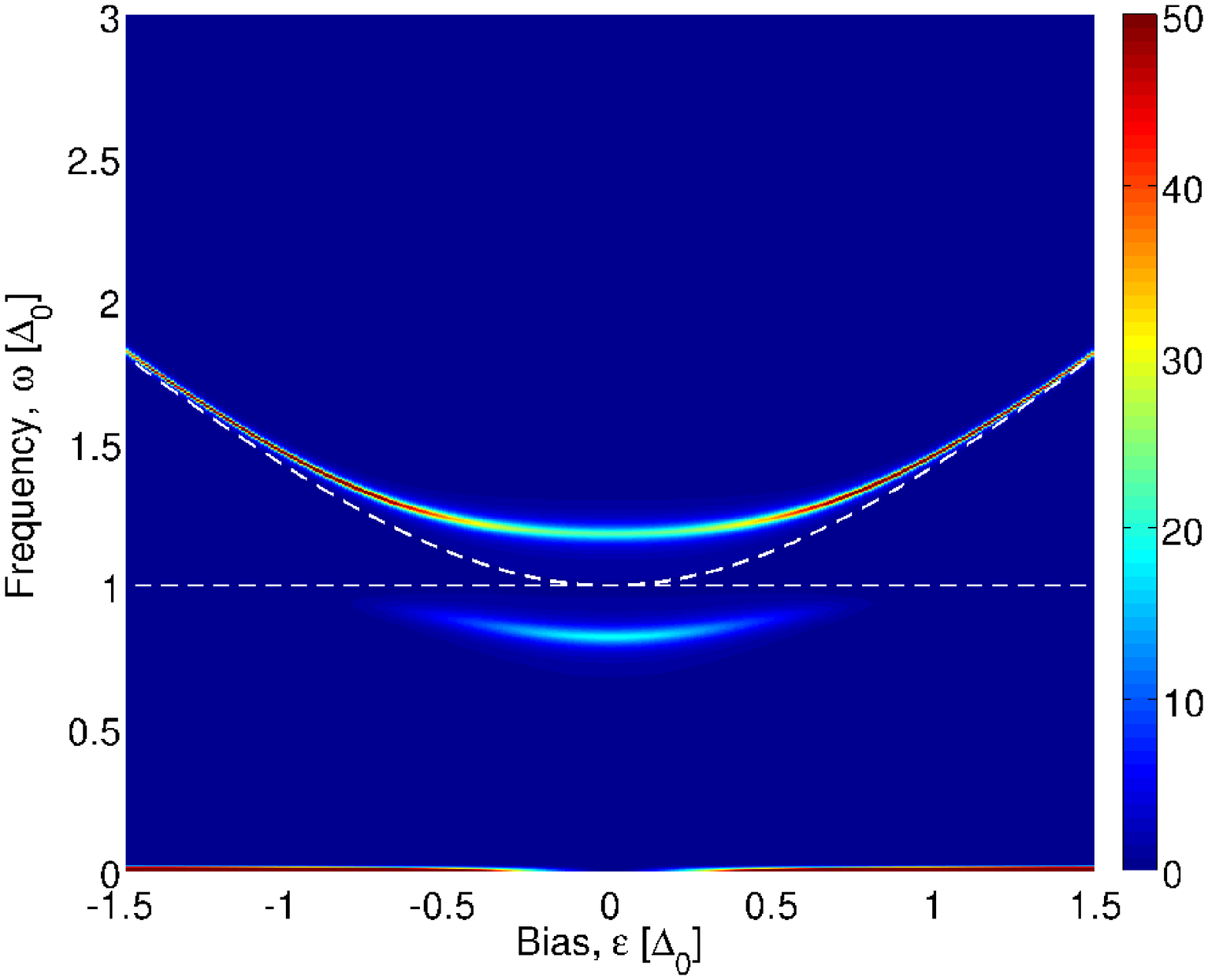}}}
\hspace{0.1 cm}
\rotatebox{-90}{
\resizebox{!}{0.47\linewidth}{
 \includegraphics{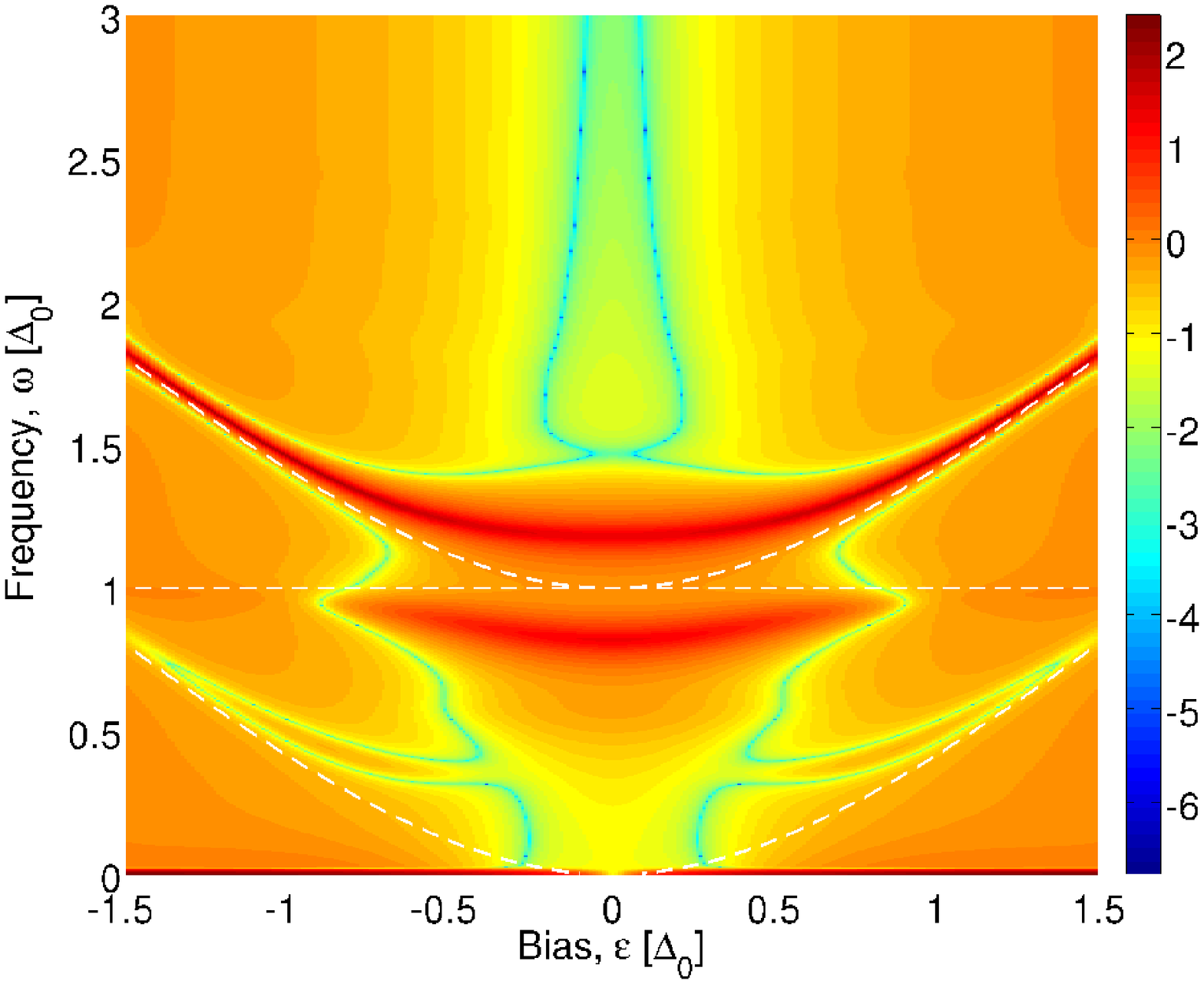}}}
\end{flushright}
\caption{Fourier transform of the symmetrized correlation function plotted versus Fourier frequency $\omega$ and qubit bias $\varepsilon$. In the left-hand graph $S(\omega)$ is plotted in linear scale, in the right-hand graph $|S(\omega)|$ in logarithmic scale. The parameters are: $\Omega =\Delta_0$, $g=0.18 \Delta_0$, $\kappa =0.0154$ and $\beta = 10 (\hbar \Delta_0)^{-1}$. The white dashed, horizontal line indicates the oscillator frequency $\omega =\Omega$. The other two dashed white lines correspond to $\omega = \Delta_{\rm b}$ and $\omega = \Delta_{\rm b} - \Omega$.  \label{FigSymmCorr}}
\end{figure}

So far we have always considered the qubit for certain values of $\varepsilon$ and finite or zero detuning. In this section, we fix the oscillator frequency at $\Omega = \Delta_0$. That means that an unbiased qubit will be at resonance with the oscillator. Changing the bias to positive or negative values will always lead to negative detuning, as $\Delta_b \geq \Delta_0$. \Fref{FigSymmCorr} shows a density plot of the Fourier transform of the symmetrized correlation function against the bias of the qubit and the Fourier frequency $\omega$. We consider this correlation function rather than $P(t)$, as it is symmetric in the bias $\varepsilon$.
The symmetrized correlation function is defined as follows \cite{Weiss}:
\begin{equation}
  S(t) = \frac{1}{2} \langle \sigma_{\rm z} (t) \sigma_{\rm z} (0) + \sigma_{\rm z} (0) \sigma_{\rm z} (t) \rangle - p_\infty^2,
\end{equation} 
where $\sigma_{\rm z} (t) = \rme^{\rmi \mathcal{H} t / \hbar} \sigma_{\rm z} \rme^{-\rmi \mathcal{H} t / \hbar} $. Expressed in terms of the population difference $P(t)$ this becomes,
\begin{equation} \label{SymmCorr}
  S(t) = P_{\rm s}(t) + p_\infty (P_{\rm a}(t)-p_\infty),
\end{equation} 
with $P_{\rm s}(t)$ and $P_{\rm a}(t)$ being symmetric and antisymmetric in $\varepsilon$ and $P(t) = P_{\rm s}(t) + P_{\rm a}(t)$. The Fourier transform of $S(t)$ is defined as
\begin{equation}
  S(\omega) = 2 \int_0^\infty dt \cos (\omega t) S(t).
\end{equation} 
Considering now \fref{FigSymmCorr} we see that for any bias the spectrum is dominated by two frequencies, namely $\omega_{10}$ and $\omega_{20}$. Detuning the system $\omega_{20}$ gets more and more important, as we could already observe in the two previous sections for the positively detuned systems. Furthermore, the peaks are shifted to higher frequency values and at $\omega = 0$ the relaxation peak occurs. We want to compare  these results to a circuit QED experiment performed by Wallraff \etal \cite{Wallraff}. There the qubit is realized by a Cooper pair box, which is coupled to a superconducting transmission line resonator. The properties of the system are determined by probing the resonator spectroscopically. The amplitude of a microwave probe beam transmitted through the resonator is measured versus the probe frequency and the gate charge of the Cooper pair box (see figure 4 in \cite{Wallraff}). Via the gate charge the qubit can be  detuned in situ from the degeneracy point. The frequency of the resonator is chosen in such a way that it is in resonance with a qubit being operated at the degeneracy point. For the resonant case two dominating frequencies, being almost equally weighted and symmetrically positioned around the cavity frequency, are observed. Going away from the degeneracy point the system becomes detuned and the frequency of the cavity dominates. The behaviour we observe in \fref{FigSymmCorr} is similar. However, as we are looking at the dynamics of the qubit, it corresponds to a spectroscopic measurement on the TLS rather than on  the oscillator. As explained above the two lowest excited states of the coupled TLS-HO system, namely $\ket{1}$ and $\ket{2}$, evolve from an almost symmetric superposition of basis states $\{ \ket{j {\rm g}}, \ket{j  \rme}  \}$ at resonance $(\varepsilon = 0)$ to the states $\ket{1 {\rm g}}$ and $\ket{0 \rm e}$ (cf the left graph in \fref{FigEnergySchemata}). For $\varepsilon=0$ the two peaks of the Rabi splitting are observed. For $\varepsilon \neq 0$, which means negative detuning in this case, the peak with the lower frequency corresponding to $\omega_{10}$ approaches more and more the frequency $\Omega$ of the oscillator, as the state $\ket{1}$ becomes $\ket{1 {\rm g}}$ for large detuning and then $\hbar \omega_{10} \approx E_{\ket{1 {\rm g}}} - E_{\ket{0 {\rm g}}} = \hbar \Omega$. Furthermore, the transition peak at $\omega_{10}$ gets weaker as also the occupation probability of $\ket{1}$ decreases. At $\varepsilon \approx \pm 0.8 \Delta_0$ the  symmetrized correlation function vanishes at $\omega_{10}$ and increases again for higher values of $|\varepsilon|$. Here, the amplitude $p_{10}$  in $P(t)$ changes its sign. In contrast the peak at $\omega_{20}$ becomes stronger with the detuning and approaches more and more the qubit splitting energy $ \hbar \Delta_{\rm b}$, as $\ket{2}$ approaches $\ket{0 \rme}$ and then $\hbar \omega_{20} \approx E_{\ket{0 \rme}} - E_{\ket{0 {\rm g}}} = \hbar \Delta_{\rm b}$. Additionally, looking at the logarithmic plot one sees around $\omega = 0.4 \Delta_0$  a peak appearing, which corresponds to the frequency $\omega_{21}$ and  is forbidden at $\varepsilon =0$. For large detuning it arises from transitions from $\ket{0 \rme}$ to $\ket{1 {\rm g}}$ and therefore has the value $\omega_{21} \approx \Delta_{\rm b} -\Omega$. The amplitude of this peak is very small compared to the peaks at $\omega_{10}$ and $\omega_{20}$ and is not resolved in the experiment of Wallraff \etal.

\section{Comparison with the Jaynes-Cummings model} \label{CompJC}
Van-Vleck perturbation theory enabled us to find approximately the eigenstates and eigenenergies of the full Hamiltonian of the TLS-HO system without performing a rotating-wave approximation. Using those eigenstates and eigenenergies in a Born-Markov master equation we could calculate the dynamics of such a system under the influence of an environmental bath. In the following we will show how the results change if we neglect counter-rotating terms in the TLS-HO Hamiltonian \eref{HQHOEn} for $\varepsilon =0$. For this we rewrite the interaction part in \eref{HQHOEn} as
\begin{equation}
  \tilde \mathcal{H}_{\rm Int} = \tilde \mathcal{H}^{\rm R}_{\rm Int} + \tilde \mathcal{H}^{\rm CR}_{\rm Int} = - \hbar g (\tilde \sigma^+ B + \tilde \sigma^- B^\dagger) - \hbar g (\tilde \sigma^- B + \tilde \sigma^+ B^\dagger),
\end{equation} 
where we identified with $ \tilde \mathcal{H}^{\rm R}_{\rm Int}$ and $\tilde \mathcal{H}^{\rm CR}_{\rm Int}$  a rotating and counter-rotating part of $\tilde \mathcal{H}_{\rm Int}$, respectively, and introduced  the two-level transition operators $\tilde \sigma^{\pm} = \frac{1}{2} (\tilde \sigma_x \pm {\rm i} \tilde \sigma_y)$. Neglecting the counter-rotating part $\tilde \mathcal{H}^{\rm CR}_{\rm Int}$ in $\tilde \mathcal{H}_{\rm QHO}$ leads to the Jaynes-Cummings Hamiltonian
\begin{equation}
 \tilde \mathcal {  H}_{\rm JC}
      = -\frac{\hbar \Delta_b}{2} \tilde \sigma_{\rm z} + \hbar \Omega B^\dagger B - \hbar g (\tilde \sigma^+ B + \tilde \sigma^- B^\dagger).
\end{equation}
This Hamiltonian can be diagonalized exactly and its eigenstate and eigenvalues can for example be found in \cite{Vogel}. In order to see the effect of not taking into account  the counter-rotating terms we diagonalize $\tilde \mathcal{H}_{\rm JC}$ using Van-Vleck perturbation theory. Looking at the formula fo the effective Hamiltonian \eref{HeffGeneral} in \ref{AppVanVleck} and keeping in mind that we set $\varepsilon$ to zero, we see that the second order contributions in $g$ vanish neglecting $\tilde \mathcal{H}^{\rm CR}_{\rm Int}$; i.e., $W_{1}$, $W_{0}$ are zero in \eref{HeffMatrix}. Considering further the transformation matrix $S$, equations  \eref{S1General} and \eref{S2General} in \ref{AppVanVleck} show that $S=0$ for $\tilde \mathcal{H}^{\rm CR}_{\rm Int}=0$. Thus, with \eref{VanVleckTrafo} we find that {\it  for the Jaynes-Cummings model $\tilde \mathcal{H}_{\rm eff}$ is identical to $\tilde \mathcal{H}_{\rm JC}$ } and therefore the eigenstates of $\tilde \mathcal{H}_{\rm eff}$ are simultaneously eigenstates of $\tilde \mathcal{H}_{\rm JC}$. Consequently, one can determine from \eref{GroundstateEn} -- \eref{Eigenenergies} the eigenstates and eigenenergies of $\tilde \mathcal{H}_{\rm JC}$. The energy of the groundstate $\ket{0}^{\rm JC} = \ket{0 g}^{\rm JC}$ is $E_{0}^{\rm JC} = - \hbar \Delta_{\rm b}/2$. For the higher states we get
\numparts
\begin{eqnarray} 
  \ket{2j+1}^{\rm JC} =\cos \left( \frac{\alpha_j^{\rm JC}}{2}  \right) \ket{(j+1){\rm g}} + \sin \left( \frac{\alpha_j^{\rm JC}}{2} \right) \ket{j {\rm e}}, \label{EigenstateJC-} \\
  \ket{2j+2}^{\rm JC} = - \sin \left( \frac{\alpha_j^{\rm JC}}{2}  \right) \ket{(j+1){\rm g}} + \cos \left( \frac{\alpha_j^{\rm JC}}{2} \right) \ket{j {\rm e}} \label{EigenstateJC+},
\end{eqnarray}
\endnumparts
corresponding to the eigenenergies 
\begin{equation} \label{EigenenergiesJC}
\fl E_{2j+1/2j+2} =  \hbar \left[ (j+\frac{1}{2}) \Omega  \mp \frac{\delta_{\rm JC}}{2 \cos \alpha_j^{\rm JC}} \right]
                             = \hbar \left[ (j+\frac{1}{2}) \Omega  \mp  \frac{1}{2} \sqrt{\delta_{\rm JC}^2+4(j+1) |\Delta|^2} \right], 
\end{equation}
with $\delta_{\rm JC} = \Delta_{\rm b} - \Omega$ and $\tan \alpha_j^{\rm JC} = 2 \sqrt{j+1} |\Delta|/\delta_{\rm JC}$. Comparing these eigenstates and eigenenergies to the ones found for $\tilde \mathcal{H}_{\rm QHO}$, \eref{VanVleckEigenstate1}, \eref{VanVleckEigenstate2} and \eref{QubitSQUIDEigenstates}, we see that the counter-rotating terms yield second order corrections in $g$ not present in the Jaynes-Cummings Hamiltonian. These corrections give rise to a very prominent effect concerning  the resonance condition between TLS and HO. From $\delta_{\rm JC}$ we find the TLS being in resonance with the oscillator for $\Omega = \Delta_b$.  Considering $\delta_j$ in \eref{deltaj} this resonance condition is shifted to 
\begin{equation} \label{BlochSiegert}
  \Omega = \Delta_{\rm b} \sqrt{1 + 2 (j+1) \frac{\Delta_0^2}{\Delta_{\rm b}^4}g^2} \approx \Delta_{\rm b} \left[ 1+ \frac{1}{2} (2j +1) \frac{\Delta_0^2}{\Delta_{\rm b}^4} g^2  + \Or (g^3) \right].
\end{equation} 
This second order correction to the resonance frequency due to counter-rotating terms is known as Bloch-Siegert shift \cite{BlochSigert}.
 The eigenstates \eref{EigenstateJC-} and \eref{EigenstateJC+} are always a superposition of two basis states of the unperturbed system. This is like for the eigenstates \eref{VanVleckEigenstate1} and \eref{VanVleckEigenstate2} of the effective Hamiltonian \eref{VanVleckTrafo}. However, in order to find the eigenstates $\ket{n}$ of $\tilde \mathcal{H}_{QHO}$ we had to apply the transformation $\exp (-\rmi S)$ on the effective eigenstates so that $\ket{n}$ is in the end a superpostion of {\it several} states of the basis $\{ \ket{j {\rm g}}; \ket{j \rme} \}$.\newline
To calculate the reduced density matrix of the qubit-HO system described by $\tilde \mathcal{H}_{\rm JC}$ and taking into account the influence of an environmental oscillator bath we can again use the Born-Markov master equation \eref{MasterEqQOBasis}. We just have to use \eref{EigenstateJC-} and \eref{EigenstateJC+} as basis states and the corresponding eigenenergies. For the population difference $P(t)$ we  apply \eref{PtQubitHO}, which becomes for $\varepsilon = 0$
\begin{eqnarray} \label{PopDiffNoBias}
 \fl P(t) = \sum_n p_{nn}(t) + \sum_{\stackrel{n,m}{n > m}} p_{nm}(t) \nonumber \\
  \fl  =  \sum_n \sum_j  2  \braket{j {\rm g}}{n} \braket{j {\rm e}}{n}  \rho_{n n}(t)   + 2  \sum_{\stackrel{n,m}{n > m}}\sum_j    \biggl[\braket{j {\rm e}}{n} \braket{m}{j {\rm g}} + \braket{j {\rm e}}{m} \braket{n}{j {\rm g}}    \biggr]  \Re \{ \rho_{nm}(t) \}.
\end{eqnarray} 

\subsection{Selection rules}
\begin{figure}[h!]
\begin{center}
   \resizebox{!}{.3 \textheight}{
  \includegraphics{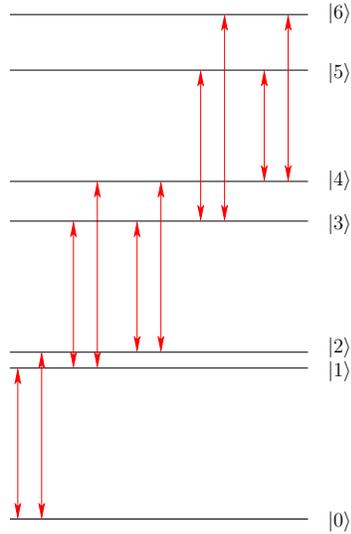}}
  \caption{Possible transitions between the first six eigenstates of the Jaynes-Cummings Hamiltonian are indicated by the red arrows. Transitions between almost degenerate levels are forbidden at zero bias ($\varepsilon = 0$). \label{FigSelecRules}}
\end{center}
\end{figure}

 Before doing a qualitative comparison between the results obtained from the original Hamiltonian $\tilde \mathcal{H}_{\rm QHO}$ and the simplified  form $\tilde \mathcal{H}_{\rm JC}$, we want to analyse which transitions between the different eigenstates of $\tilde \mathcal{H}_{\rm JC}$ yield contributions to $P(t)$. For this it is helpful to rewrite \eref{EigenstateJC-} and \eref{EigenstateJC+}. In the following we neglect the upper index JC denoting an eigenstate of $\tilde \mathcal{H}_{\rm JC}$. For the state $\ket{n}$ we have three possibilities: the first one corresponds to $n=0$. In this case the only non-vanishing component of $\ket{0}$ is $\braket{0 g}{0}$. Second, $n$ can be an even number, which means, expressed in terms of oscillator quanta, that $n = 2j+2$. Then
\begin{equation}
  \ket{n_{\rm ev}} = -\sin (\alpha_{\frac{n-2}{2}}/2 ) \ket{\left(\frac{n}{2}\right) {\rm g}} + \cos ( \alpha_{\frac{n-2}{2}} /2 ) \ket{\left( \frac{n-2}{2} \right) {\rm e}}. \label{EvenState}
\end{equation} 
And third for an odd state $n=2j+1$ we find 
\begin{equation}
  \ket{n_{\rm od}} = \cos ( \alpha_{\frac{n-1}{2}} /2)  \ket{\left(\frac{n+1}{2}\right) {\rm g}} + \sin ( \alpha_{\frac{n-1}{2}} /2 ) \ket{\left(\frac{n -1}{2}\right)  {\rm e}}. \label{OddState}
\end{equation} 
It is quite easy to see that the part $p_{nn}(t)$ in \eref{PopDiffNoBias} vanishes for any $n$. That means that the diagonal elements of the reduced density matrix yield no contributions to $P(t)$ for $\varepsilon = 0$ and the equilibrium value of the dynamics will be also zero. Not so easy to see are the combinations of $n$ and $m$ yielding contributions to the off-diagonal part $p_{nm}(t)$.
Considering transitions to the groundstate ($m=0$), we find using that $\braket{0 g}{0} = 1$ that
\begin{equation}
   p_{n_{\rm ev}0} = 2 \braket{0 {\rm e}}{n_{\rm ev}}  \Re \{ \rho_{n_{\rm ev} 0}(t) \} \quad \textrm{and} \quad p_{n_{\rm od}0} = 2 \braket{0 {\rm e}}{n_{\rm od}}  \Re \{ \rho_{n_{\rm od} 0}(t) \}.
\end{equation}
With (\ref{EvenState}) and (\ref{OddState}) these elements are non-zero only if $n=1$ or $n=2$.  For the more general case with $n \neq m$ and both being different from zero we have to investigate products of components like $\braket{j {\rm e}}{n} \braket{m}{j {\rm g}}$. For $p_{n_{\rm ev} m_{\rm ev}}(t)$ we find that
\begin{equation}
  \braket{j {\rm e}}{n_{\rm ev}} \braket{m_{\rm ev}}{j {\rm g}} \neq 0 \quad \textrm{if} \quad j=\frac{n_{\rm ev}-2}{2} \quad \textrm{and} \quad j=\frac{m_{\rm ev}}{2} 
\end{equation} 
and that 
\begin{equation}
   \braket{j {\rm e}}{m_{\rm ev}} \braket{n_{\rm ev}}{j {\rm g}} \neq 0 \quad \textrm{if} \quad j=\frac{m_{\rm ev}-2}{2} \quad \textrm{and} \quad j=\frac{n_{\rm ev}}{2}, 
\end{equation}
so that 
\begin{equation}
 p_{n_{\rm ev} m_{\rm ev}}(t) \neq 0 \quad \textrm{if} \quad |n_{\rm ev} - m_{\rm ev} | = 2.
\end{equation} 
For the case of transitions between odd states, one gets that
\begin{equation}
  \braket{j {\rm e}}{n_{\rm od}} \braket{m_{\rm od}}{j {\rm g}} \neq 0 \quad \textrm{if} \quad j=\frac{n_{\rm od}-1}{2} \quad \textrm{and} \quad j=\frac{m_{\rm od} +1}{2} 
\end{equation} 
and similarly under exchange of $n_{\rm od}$ and $m_{\rm od}$, so that
\begin{equation}
 p_{n_{\rm od} m_{\rm od}}(t) \neq 0 \quad \textrm{if} \quad |n_{\rm od} - m_{\rm od} | = 2.
\end{equation}
For transitions from an even to an odd state and vice versa we have to pay attention to the fact that 
\begin{equation}
  \braket{j {\rm e}}{n_{\rm ev}} \braket{m_{\rm od}}{j {\rm g}} \neq 0 \quad \textrm{if} \quad j=\frac{n_{\rm ev}-2}{2} \quad \textrm{and} \quad j=\frac{m_{\rm od} + 1}{2}, 
\end{equation} 
which yields the selection rule $n_{\rm ev} - m_{\rm od}  = 3$. Further,
\begin{equation}
   \braket{j {\rm e}}{m_{\rm od}} \braket{n_{\rm ev}}{j {\rm g}} \neq 0 \quad \textrm{if} \quad j=\frac{m_{\rm od}-1}{2} \quad \textrm{and} \quad j=\frac{n_{\rm ev}}{2}, 
\end{equation}
yielding $m_{\rm od} - n_{\rm ev}  = 1$. To sum up: $p_{nm}(t)$ is non-zero if one of the three following cases is valid:
  $|n_{\rm ev} - m_{\rm ev} | = 2$, $|n_{\rm od} - m_{\rm od} | = 2$, $n_{\rm ev} - m_{\rm od}  = 3$ or $m_{\rm od} - n_{\rm ev}  = 1$.
The allowed transitions are shown in figure \ref{FigSelecRules}. 
One sees that transitions between almost degenerate levels are forbidden. This behaviour we have also found in \sref{LowTempApprox} for the non-dissipative dynamics resulting from $\tilde \mathcal{H}_{\rm QHO}$.

\subsection{Comparison of the two models}
\begin{figure}[h]
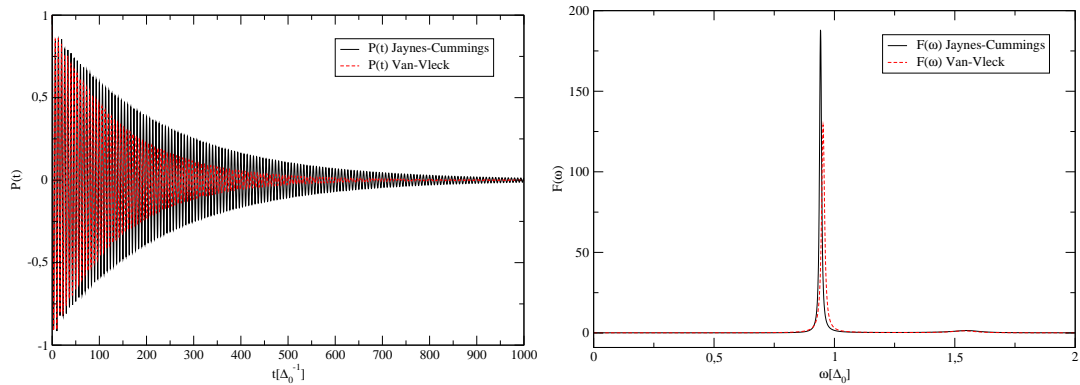

 \begin{center}
  \resizebox{.45\textwidth}{!}{
    \includegraphics{Figures/figure15a.eps}}
  \resizebox{.45\textwidth}{!}{
    \includegraphics{Figures/figure15b.eps}}
  \caption{Dynamics $P(t)$ of the population difference and its  Fourier transform $F(\omega)$ for positive detuning ($\Omega = 1.5 \Delta_0$). The parameters are in units of $\Delta_0$: $\varepsilon=0$, $g=0.18$, $\kappa=0.0154$ and $\beta=10$. The red dashed line shows the solution obtained numerically from the master equation using the eigenstates and eigenenergies from the full Hamiltonian $\mathcal{H}_{\rm QHO}$. The solid line shows the results obtained from the Jaynes-Cummings Hamiltonian. \label{PositiveDetuning}}
\end{center}
\end{figure}

\begin{figure}[h]
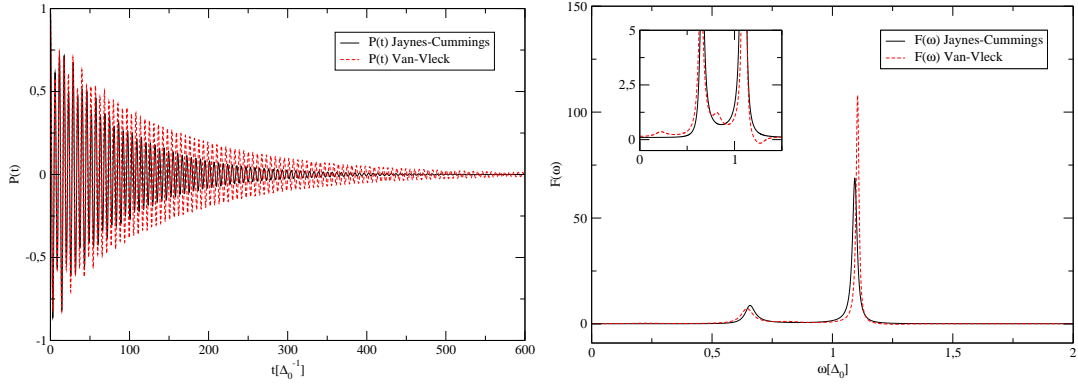

 \begin{center}
  \resizebox{.45\textwidth}{!}{
    \includegraphics{Figures/figure16a.eps}}
  \resizebox{.45\textwidth}{!}{
    \includegraphics{Figures/figure16b.eps}}
 \caption{Dynamics $P(t)$ of the population difference and its  Fourier transform $F(\omega)$ for negative detuning ($\Omega = 0.75 \Delta_0$). The remaining parameters are the same as in figure \ref{PositiveDetuning}. The red dashed line shows the solution obtained numerically from the master equation using the eigenstates and eigenenergies from the full Hamiltonian $\mathcal{H}_{\rm QHO}$. The solid line shows the results obtained from the Jaynes-Cummings Hamiltonian. \label{NegativeDetuning}}
\end{center}
\end{figure}

\begin{figure}[h!]
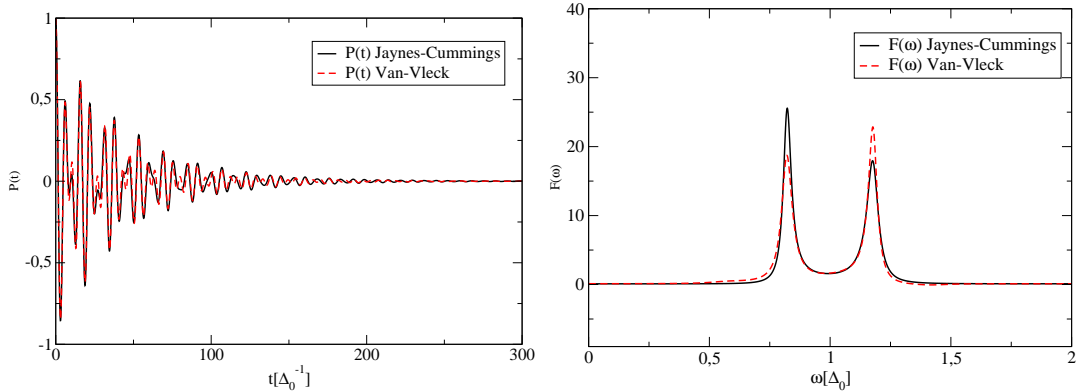

 \begin{center}
  \resizebox{.45\textwidth}{!}{
    \includegraphics{Figures/figure17a.eps}}
  \resizebox{.45\textwidth}{!}{
    \includegraphics{Figures/figure17b.eps}}
 \caption{Dynamics $P(t)$ of the population difference and its  Fourier transform $F(\omega)$ for the resonant case ($\Omega = \Delta_0$). The remaining parameters are the same as in figure \ref{PositiveDetuning}. The red dashed line shows the solution obtained numerically from the master equation using the eigenstates and eigenenergies from the full Hamiltonian $\mathcal{H}_{\rm QHO}$. The solid line shows the results obtained from the Jaynes-Cummings Hamiltonian. \label{NoDetuning}}
\end{center}
\end{figure}

In the following we compare the numerical solution of the Born-Markov master equation originating from the  Jaynes-Cummings Hamiltonian with the solution using the eigenstates of the full Hamiltonian $\tilde \mathcal{H}_{\rm QHO}$, which we found by applying Van-Vleck perturbation theory. The fixed parameters we use are in units of $\Delta_0$: $\varepsilon =0$, $g=0.18$, $\kappa = 0.0154$ and $\beta = 10$. The oscillator frequency $\Omega$ is varied. For all the three possible cases (positive, negative and zero detuning) one notices from figures \ref{PositiveDetuning} -- \ref{NoDetuning}  that the Jaynes-Cummings approach {\it underestimates} the dephasing rate $\Gamma_{10}$ (means a larger peak at frequency $\omega_{10}$) and  {\it overestimates} the rate $\Gamma_{20}$ (smaller peak at $\omega_{20}$) compared to the approach with the full Hamiltonian.\newline
The case of positive detuning ($\Omega > \Delta_0$) is shown in figure \ref{PositiveDetuning}. As here the dynamics is dominated by the frequency $\omega_{10}$,  the equilibrium value is reached on a too long time scale using the Jaynes-Cummings Hamiltonian.
On the contrary, for negative detuning ($\Omega < \Delta_0$), which is shown in figure \ref{NegativeDetuning} and where $\omega_{20}$ dominates, the equilibrium value is reached too fast within the Jaynes-Cummings approach. Furthermore, considering the graph of the Fourier transform we find that small contributions which come from higher level transitions, and which have already been discussed in \sref{UnbiasedQubit}, are not caught be the Jaynes-Cummings approach.
For the resonant case ($\Omega = \Delta_0$) we find that the Jaynes-Cummings method predicts $\omega_{10}$ to be slightly dominating whereas the approach starting from $\mathcal{H}_{\rm QHO}$ results in $\omega_{20}$ being dominating.
The reason for this discrepancy is that due to the counter-rotating terms we have for $\mathcal{H}_{\rm QHO}$ no symmetric or antisymmetric superposition of the unperturbed eigenstates at $\Omega = \Delta_0$ in contrast to the Jaynes-Cummings model. \newline
To conclude this section we can say that for an unbiased TLS-HO system the Jaynes-Cummings model gives a good insight in the qualitative behaviour of $P(t)$ both for a slightly detuned and a non-detuned system. However, it under- or overestimates dephasing times for the system. Furthermore, we find taking into account counter-rotating terms in $\tilde \mathcal{H}_{QHO}$ that at $\Omega = \Delta_{\rm b}$ the dressed eigenstates are not a symmetric or antisymmetric superposition of the uncoupled states.  Moreover, the effects of transitions between states of different manifolds are neglected.

\section{Conclusions}
In conclusion, we discussed the dynamics of a biased and unbiased TLS coupled through a harmonic oscillator to an environmental bath described by an Ohmic spectral density. In particular, we examined the regime of weak damping and moderate coupling between oscillator and TLS. An equivalent description of our system is provided by the spin-boson model with a structured spectral density. In contrast to many other works in this field, our starting point was not the Jaynes-Cummings Hamiltonian, but a more general one given in  \eref{StartingHamiltonian}, where no initial rotating wave approximation has been applied. In \sref{PopulationDifference} we provided with \eref{PtQubitHO} a formal expression for the population difference and showed in \sref{SecEnSpec} how the Hamiltonian of the coupled qubit-HO system can be diagonalized approximately using Van-Vleck perturbation theory. This approach is valid both for the oscillator being in resonance with the qubit ($\Omega = \Delta_{\rm b}$) and for finite detuning ($|\delta| \neq 0$). In \sref{DynamicsNoDiss} an analytical expression  for the non-dissipative dynamics was provided up to second order in the qubit-HO coupling $g$ taking into account the infinite Hilbert space of the system. For low temperatures ($k_{\rm B} T < \hbar \Omega, \hbar \Delta_{\rm b} $) we truncated the Hilbert space and found that transition processes between the groundstate and the two first excited energy levels of the qubit-HO system dominate the dynamics (\sref{LowTempApprox}). In \sref{InfluenceEnvironment} the influence of the bath was taken into account by solving the Born-Markov master equation  for the density matrix of the qubit-HO system. To do this analytically we considered two variants of the secular approximation: first, in \sref{FSA}, the full secular approximation (FSA), where all fast oscillating terms are neglected, and second, in \sref{PSA}, the partial secular approximation, where  attention  was paid to the fact that the first two excited energy levels are almost degenerate.  Using an ansatz for the long time dynamics in \sref{AnsatzFSA}, we could provide a general expression for the relaxation and dephasing rates of the qubit, showing that the relaxation time can be enhanced by detuning the oscillator into the off-resonant regime. It was found that all three approaches agree quite well with the numerical solution of the master equation. The dynamics of  both a biased and an unbiased qubit were intesively studied for zero and finite detuning in \sref{DiscussionQubits}. The results agree qualitatively with the numerical findings within the ab-initio QUAPI approach \cite{Thorwart}.  Furthermore, in \sref{UnbiasedQubit} a good agreement with the results of the weak damping approximation performed in \cite{Nesi} for a symmetric spin-boson model was found. Besides, we saw that at resonance ($\Omega = \Delta_{\rm b}$) the first two excited qubit oscillator states are not a symmetric or antisymmetric superposition of the states $\ket{0 {\rm e}}$ and $\ket{1 {\rm g}}$, as predicted by the Jaynes-Cummings model, and thus could give an explanation for the differently weighted peaks of $\omega_{10}$ and $\omega_{20}$ in the Fourier spectrum of the dynamics. We further could explain the dominance of frequency $\omega_{20}$ in the case of negative detuning $(\Omega<\Delta_{\rm b})$ and of frequency $\omega_{10}$ for positive detuning, respectively. Moreover, we showed that for large negative detuning $\omega_{20}$ approaches the energy splitting $\Delta_{\rm b}$ of the qubit, whereas $\omega_{10}$ approximates the oscillator frequency $\Omega$. This behaviour agrees nicely with spectroscopic experiments performed on a circuit QED architecture \cite{Wallraff}.
In \sref{CompJC} we compared our results for an unbiased system to the ones obtained starting with the Jaynes-Cummings model. We visualize the effects of the counter-rotating terms in the Bloch-Siegert shift of the resonance frequency and in contributions of states with larger oscillator number to the TLS dynamics. Apart from this  the Jaynes-Cummings model and our approach agree quite well for the non-dissipative case. Also for the dissipative case at resonance an initial RWA represents a good approximation to our starting Hamiltonian and seems to be favourable as it is analytically exactly diagonalizable.  For  detuned systems, however, we find discrepancies concerning relaxation and dephasing times and a RWA becomes  less appropriate to give precise results. Thus, we think that our approach represents an improvement as it is valid in a wider parameter range avoiding an initial rotating wave approximation. To our knowledge it provides for the first time analytical results for the dynamics of an unbiased {\it and} biased qubit coupled to a structured environment being valid both in the resonant {\it and} off-resonant regime. Furthermore, due to the generality of the qubit-oscillator model, we expect our results to be of interest for a wide range of experimental applications.

\ack
We acknowledge financial support under DFG program SFB631. Further we would like to thank G. Begemann for his help with some of the figures.

\appendix

\section{Formula for the dynamics} \label{AppPt}
Here, we show how to derive  \eref{PtQubitHO} given in \sref{PopulationDifference}. In order to trace out the oscillator degrees of freedom we transform $\rho_{nm}(t)$ into the basis $\{ \ket{j {\rm g}} ; \ket{j {\rm e}}  \}$ and by using $\rho_{nm} = \rho_{mn}^*$ we find
\numparts
\begin{eqnarray}
  \fl \rho_{j {\rm g},j {\rm g}} (t) = \bra{j {\rm g}} \rho(t) \ket{j {\rm g}} = \sum_n \braket{j {\rm g}}{n}^2 \rho_{nn}(t) + \sum_{\stackrel{n,m}{n \neq m}} \Re \{\rho_{nm}(t) \} \braket{j {\rm g}}{n} \braket{m}{j {\rm g}}\\
 \fl \rho_{j {\rm e},j {\rm e}} (t) = \bra{j {\rm e}} \rho (t) \ket{j {\rm e}} = \sum_n \braket{j {\rm e}}{n}^2 \rho_{nn} (t) + \sum_{\stackrel{n,m}{n \neq m}} \Re \{\rho_{nm} (t) \} \braket{j {\rm e}}{n} \braket{m}{j {\rm e}}.
\end{eqnarray}
\endnumparts
Performing the trace over the oscillator 
\numparts
\begin{eqnarray}
 \fl  \rho_{{\rm red;gg}} (t) = \bra{{\rm g}} \rho_{\rm red} (t) \ket{{\rm g}} = \sum_{j=0}^\infty \sum_n \braket{j {\rm g}}{n}^2 \rho_{nn} (t) +\sum_{j=0}^\infty \sum_{\stackrel{n,m}{n \neq m}} \Re \{\rho_{nm}(t) \} \braket{j {\rm g}}{n} \braket{m}{j {\rm g}}, \\
 \fl  \rho_{\rm red;ee} (t) = \bra{{\rm e}} \rho_{\rm red} (t) \ket{{\rm e}} = \sum_{j=0}^\infty \sum_n \braket{j {\rm e}}{n}^2 \rho_{nn} (t) +\sum_{j=0}^\infty \sum_{\stackrel{n,m}{n \neq m}} \Re \{\rho_{nm}(t) \} \braket{j {\rm e}}{n} \braket{m}{j {\rm e}}.
\end{eqnarray}
\endnumparts
 Similarily, we find for the off-diagonal elements of the reduced density matrix
\numparts
 \begin{eqnarray}
  \fl \rho_{\rm red;eg} (t) = \sum_{j=0}^\infty \sum_n \braket{j {\rm e}}{n} \rho_{nn}(t) \braket{n}{j {\rm g}} \nonumber \nolinebreak\\
      + \frac{1}{2} \sum_{j=0}^\infty \sum_{\stackrel{n,m}{n \neq m}} \left[ \braket{j {\rm e}}{n} \rho_{nm}(t) \braket{m}{j {\rm g}} + \braket{j {\rm e}}{m} \rho_{nm}^*(t) \braket{j {\rm g}}{n} \right], 
\end{eqnarray}
\begin{eqnarray}
  \fl \rho_{\rm red;ge} (t) = \sum_{j=0}^\infty \sum_n \braket{j {\rm g}}{n} \rho_{nn}(t) \braket{n}{j {\rm e}} \nonumber \\
       + \frac{1}{2} \sum_{j=0}^\infty \sum_{\stackrel{n,m}{n \neq m}} \left[ \braket{j {\rm g}}{n} \rho_{nm}(t) \braket{m}{j {\rm e}} + \braket{j {\rm g}}{m} \rho_{nm}^*(t) \braket{j {\rm e}}{n} \right].
\end{eqnarray}
\endnumparts
Using \eref{LocaltoEn1} and \eref{LocaltoEn2} in \eref{dynamics} we can express $P(t)$ in the energy basis, yielding
\begin{equation} \label{PtTLS}
  \eqalign{
  \fl P(t) = \cos \Theta \left[ \rho_{\rm red;gg}(t) - \rho_{\rm red;ee}(t) \right] + \sin \Theta \left[ \rho_{\rm red;ge}(t)+\rho_{\rm red;eg}(t) \right]\\
   \fl = \cos(\Theta) \biggl(    \sum_{j=0}^\infty \sum_n \left[ \braket{j{\rm g}}{n}^2 - \braket{j {\rm e}}{n}^2 \right] \rho_{nn} (t) \\
    +\sum_{j=0}^\infty \sum_{\stackrel{n,m}{n \neq m}} \left[ \braket{j {\rm g}}{n} \braket{m}{j {\rm g}} -  \braket{j {\rm g}}{n} \braket{m}{j {\rm g}} \right] \Re \{\rho_{nm}(t) \} \biggr) \\
    \fl \phantom{\mathrel{=}} + \sin (\Theta) \biggl( 2 \sum_{j=0}^\infty \sum_n \braket{j {\rm e}}{n} \rho_{nn}(t) \braket{n}{j {\rm g}} \\
           + \sum_{j=0}^\infty \sum_{\stackrel{n,m}{n \neq m}} \left[ \braket{j {\rm e}}{n}\braket{m}{j {\rm g}} + \braket{j {\rm e}}{m}\braket{j {\rm g}}{n}\right] \Re \{\rho_{nm}(t)\}\biggr)}.
\end{equation}
With \eref{DensMatNoDiss} we can write
\begin{equation} 
  P(t) = \sum_n p_{nn}(t) + \sum_{\stackrel{n,m}{n > m}} p_{nm}(t) \cos{\omega_{nm} t},
\end{equation}  
where
\numparts
\begin{eqnarray}
  \fl  p_{nn}(t) =  \sum_j \left\{ \cos \Theta \biggl[ \braket{j {\rm g}}{n}^2 - \braket{j {\rm e}}{n}^2 \biggr]  + 2 \sin \Theta \braket{j {\rm g}}{n} \braket{j {\rm e}}{n} \right\} \rho_{n n}(t), \\
  \fl p_{nm}(t) = 2 \sum_j \biggl\{ \cos \Theta \biggl[\braket{j {\rm g}}{n} \braket{m}{j {\rm g}} - \braket{j {\rm e}}{n} \braket{m}{j {\rm e}}    \biggr] \nonumber\\
     + \sin \Theta \biggl[\braket{j {\rm e}}{n} \braket{m}{j {\rm g}} + \braket{j {\rm e}}{m} \braket{n}{j {\rm g}}    \biggr] \biggr\} \Re \{\rho_{nm}(t) \}. 
\end{eqnarray}
\endnumparts

\section{Van-Vleck perturbation theory}\label{AppVanVleck}
Let us consider the Hamiltonian
\begin{equation}
  \mathcal{H} = \mathcal{H}_0+V
\end{equation} 
consisting of the free Hamiltonian $\mathcal{H}_0$ and a small perturbation $V \sim g$, which is proportional to the coupling constant $g$. Additionally we assume, that the energy levels $E_{j,\alpha}$ of $\mathcal{H}_0$ are grouped into manifolds, with $\alpha$ being the index of the manifold and $i$ is used to distinguish between different energy levels within the same manifold. The energy levels $E_{j,\alpha}$ are eigenenergies of $\mathcal{H}_0$:
\begin{equation}
 \mathcal{H}_0 \ket{j,\alpha} = E_{j,\alpha} \ket{j, \alpha}.
\end{equation}
Through the perturbation $V$ different manifolds are coupled together. As long as the coupling $g$ is small, namely that
$|\bra{j,\alpha} V \ket{j,\beta}| \ll |E_{j,\alpha}-E_{j,\beta}|$ for $\alpha \neq \beta$, 
also the energy levels of the total Hamiltonian $\mathcal{H}$ are clustered into manifolds. Using the transformation
 $\mathcal{H}_{\rm eff} = \rme^{\rmi S} \mathcal{H} \rme^{-\rmi S}$,
we construct an effective Hamiltonian $\mathcal{H}_{\rm eff}$, which acts only within the individual manifolds; i.e.,
 $\bra{j,\alpha} \mathcal{H}_{\rm eff} \ket{j,\beta} = 0$  for $\alpha \neq \beta$, 
and has the same eigenvalues as  $\mathcal{H}$ within the manifolds.
We expand $S$ and $\mathcal{H}_{\rm eff}$ in terms of the small parameter $g$ up to second order:
\begin{equation}
 \fl S= S^{(1)}+S^{(2)}+ \Or(g^3) \quad {\rm and} \quad \mathcal{H}_{\rm eff}=\mathcal{H}_{\rm eff}^{(0)} + \mathcal{H}_{\rm eff}^{(1)} + \mathcal{H}_{\rm eff}^{(2)} + \Or(g^3)
\end{equation} 
For calculating $S^{(1/2)}$ and $\mathcal{H}_{\rm eff}^{(1/2)}$ we use that
$  \bra{j,\alpha} \mathcal{H}_{eff}^{(1/2)} \ket{j,\beta} =0$ for $\alpha \neq \beta$
and furthermore choose that $S$ has no matrix elements within a manifold, namely
 $\bra{j,\alpha} iS^{(1/2)} \ket{j,\alpha} = 0$. 
Now, one can iteratively calculate $S$ and $\mathcal{H}_{\rm eff}$ order by order. For details see \cite{Cohen1992}. Here, we give only the results. For the transformation one has
\begin{equation} \label{S1General}
  \bra{j,\alpha} iS^{(1)} \ket{j,\beta} = \frac{  \bra{j,\alpha} V \ket{j,\beta}}{E_{j,\alpha} - E_{j,\beta}}, \textrm{\quad for \quad} \alpha \neq \beta,
\end{equation}
and
\begin{eqnarray} \label{S2General}
  \fl \bra{j,\alpha} iS^{(2)} \ket{j,\beta} = \frac{1}{2} \sum_{k,\gamma \neq \alpha, \beta} \frac{\bra{j,\alpha} V \ket{k,\gamma} \bra{k,\gamma} V \ket{j,\beta}}{E_{j, \beta}-E_{j, \alpha}} \left[ \frac{1}{E_{k,\gamma}- E_{j, \alpha}}+ \frac{1}{E_{k,\gamma} - E_{j, \beta} }   \right] \nonumber \\
        + \sum_k \frac{1}{E_{j,\beta} - E_{j,\alpha}} \frac{\bra{j,\alpha} V \ket{k,\beta} \bra{k,\beta} V \ket{j,\beta}}{E_{k, \beta}-E_{j, \alpha}} \nonumber \\
        + \sum_k \frac{1}{E_{j,\beta} - E_{j,\alpha}} \frac{\bra{j,\alpha} V \ket{k,\alpha} \bra{k,\alpha} V \ket{j,\beta}}{E_{k, \alpha}-E_{j, \beta}}, \textrm{\quad for \quad} \alpha \neq \beta.
\end{eqnarray}
The effective Hamiltonian is up to second order
\begin{eqnarray} \label{HeffGeneral}
  \fl \bra{i,\alpha} \mathcal{H}_{eff}  \ket{j, \alpha} = E_{j,\alpha} \delta_{ij} + \bra{i, \alpha}V\ket{j,\alpha}  \\
                 \hspace{-1.5 cm} + \frac{1}{2} \sum_{k, \gamma \neq \alpha} \bra{i, \alpha}V\ket{k,\gamma} \bra{k, \gamma}V\ket{j,\alpha} \left[ \frac{1}{E_{i,\alpha} - E_{k,\gamma}} + \frac{1}{E_{j,\alpha} - E_{k,\gamma}} \right] + \Or(g^3).\nonumber
\end{eqnarray}

In the case of the Hamiltonian $\tilde \mathcal {H}_{\rm QHO}$ the first order matrix elements are
\numparts
\begin{eqnarray} \label{SFirstOrder}
  \rmi S^{(1)}_{{\rm e}_{j-1}{\rm e}_j} &= \sqrt{j} \frac{\varepsilon}{\Delta_b \Omega} g,\\
  \rmi S^{(1)}_{{\rm g}_j{\rm g}_{j+1}} &= -\sqrt{j+1} \frac{ \varepsilon}{\Delta_b \Omega} g,\\
  \rmi S^{(1)}_{{\rm g}_j{\rm e}_{j+1}} &= \sqrt{j+1} \frac{ \Delta_0}{\Delta_b (\Delta_b +\Omega)} g,
\end{eqnarray}
\endnumparts
and for the second order contributions
\numparts
\begin{eqnarray} 
   \rmi S^{(2)}_{{\rm e}_j{\rm g}_{j+2}} &= 2 \sqrt{(j+1) (j+2)} \frac{\varepsilon \Delta_0}{\Delta_b^2 \Omega (2\Omega-\Delta_b)} g^2 \label{SSecondOrder1},\\
   \rmi S^{(2)}_{{\rm e}_j{\rm e}_{j+2}} &= - \sqrt{(j+1)(j+2)} \frac{\Delta_0^2}{2\Delta_b^2 \Omega (\Delta_b + \Omega)} g^2,\\
   \rmi S^{(2)}_{{\rm g}_j{\rm e}_j} &= - \frac{1}{2} (2j+1) \frac{\varepsilon \Delta_0}{\Delta_b^2 \Omega (\Delta_b + \Omega)} g^2, \\
   \rmi S^{(2)}_{{\rm g}_j{\rm g}_{j+2}} &= \frac{1}{2} \sqrt{(j+1)(j+2)} \frac{\Delta_0^2}{\Delta_b^2 \Omega (\Omega + \Delta_b)} g^2, \\
   \rmi S^{(2)}_{{\rm g}_j{\rm e}_{j+2}} &= - \sqrt{(j+1)(j+2)} \frac{\varepsilon \Delta_0}{\Delta_b^2 \Omega (\Delta_b + \Omega)(\Delta_b + 2 \Omega)} g^2, \label{SSecondOrder5}
\end{eqnarray}
\endnumparts
where e. g. 
\begin{equation}
  S^{(1)}_{{\rm e}_{j-1}{\rm e}_j} =\bra{(j-1) {\rm e}} S \ket{j {\rm e}}.
\end{equation}
These matrix elements change sign under index transposition and all other matrix elements vanish. Finally, we get  the transformation up to second order in $g$:
\begin{equation}
  e^{\pm \rmi S} = \mathds{1} \pm \rmi S^{(1)} \pm \rmi S^{(2)} + \frac{1}{2} \rmi S^{(1)} \rmi S^{(1)} + \Or (g^3). \label{FullTrafo} \\
\end{equation}

\section{Oscillator matrix elements}\label{AppOscillatorMatrixElements}
Here, we give the matrix elements $X_{nm}$ specified in \sref{OscillatorMatrixElements}.
\begin{eqnarray*}
  \fl X_{2j+1, 2j+1} = - 2  L_0 \cos \alpha_j  + \sqrt{j+1}  L_{\rm q,osc}^- \sin \alpha_j, \\
  \fl X_{2j+1, 2j+2} = 2  L_0 \sin \alpha_j + \sqrt{j+1}  L_{\rm q,osc}^- \cos \alpha_j,\\
  \eqalign{
  \fl X_{2j+1, 2j+3} =  L_{\rm q} \cos(\alpha_j/2) \sin(\alpha_{j+1}/2) + \sqrt{j+2} (1+L_{\rm osc}) \cos(\alpha_j/2) \cos(\alpha_{j+1}/2) \\ 
   + \sqrt{j+1} (1-L_{\rm osc}) \sin(\alpha_j/2) \sin(\alpha_{j+1}/2)},\\
  \eqalign{
  \fl X_{2j+1, 2j+4} =  L_{\rm q} \cos(\alpha_j/2) \cos(\alpha_{j+1}/2) - \sqrt{j+2} (1+L_{\rm osc}) \cos(\alpha_j/2) \sin(\alpha_{j+1}/2)\\
      + \sqrt{j+1} (1-L_{\rm osc}) \sin(\alpha_j/2) \cos(\alpha_{j+1}/2)},\\
  \fl X_{2j+1, 2j+5} = \sqrt{j+2}  L_{\rm q,osc}^+ \cos (\alpha_j/2) \sin(\alpha_{j+2}/2),\\
  \fl X_{2j+1, 2j+6} = \sqrt{j+2}  L_{\rm q,osc}^+ \cos (\alpha_j/2) \cos(\alpha_{j+2}/2),\\
  \fl X_{2j+2, 2j+2} =  2  L_0 \cos \alpha_j  - \sqrt{j+1}  L_{\rm q,osc}^- \sin \alpha_j, \\
  \eqalign{
  \fl X_{2j+2, 2j+3} =  - L_{\rm q} \sin(\alpha_j/2) \sin(\alpha_{j+1}/2) - \sqrt{j+2} (1+L_{\rm osc}) \sin(\alpha_j/2) \cos(\alpha_{j+1}/2) \\
     + \sqrt{j+1} (1-L_{\rm osc}) \cos(\alpha_j/2) \sin(\alpha_{j+1}/2)},\\
  \eqalign{
  \fl X_{2j+2, 2j+4} =  - L_{\rm q} \sin(\alpha_j/2) \cos(\alpha_{j+1}/2) + \sqrt{j+2} (1+L_{\rm osc}) \sin(\alpha_j/2) \sin(\alpha_{j+1}/2)\\
      + \sqrt{j+1} (1-L_{\rm osc}) \cos(\alpha_j/2) \cos(\alpha_{j+1}/2)},\\
  \fl X_{2j+2, 2j+5} = - \sqrt{j+2}  L_{\rm q,osc}^+ \sin(\alpha_j/2) \sin(\alpha_{j+2}/2),\\
  \fl X_{2j+2, 2j+6} = - \sqrt{j+2}  L_{\rm q,osc}^+ \sin (\alpha_j/2) \cos(\alpha_{j+2}/2).
\end{eqnarray*}
Matrix elements including the groundstate are given separately because of the special shape of $\ket{0}$:
\begin{eqnarray*}
  X_{0,0} = - 2 L_0,\\
  X_{0,1} = \sin (\alpha_0/2) L_{\rm q} + \cos (\alpha_0/2) (1 + L_{\rm osc} ), \\
  X_{0,2} = \cos (\alpha_0/2) L_{\rm q} - \sin (\alpha_0/2) (1 + L_{\rm osc} ), \\
  X_{0,3} = \sin (\alpha_1/2) L_{\rm q, osc}^+,\\
  X_{0,4} = \cos (\alpha_1/2) L_{\rm q, osc}^+.
\end{eqnarray*}
All other matrix elements are zero.

\section{Rate coefficients for the off-diagonal density matrix elements} \label{RateCoeff}
Here, we give the rate coefficients of the master equation for the reduced density matrix elements. They are:
\begin{equation}
  \fl \mathcal{L}_{01,01} = \frac{4 \kappa}{\hbar \beta} L_0^2 (2 \cos \alpha_0 -\cos^2 \alpha_0 -1) - \frac{1}{2} \mathcal{L}_{00,11}, \label{L0101}
\end{equation} 
\begin{equation}
  \fl \mathcal{L}_{02,02} = \frac{4 \kappa}{\hbar \beta} L_0^2 (-2 \cos \alpha_0 -\cos^2 \alpha_0 -1) - \frac{1}{2} \mathcal{L}_{00,22}, \label{L0202}
\end{equation}
\begin{equation}
  \fl \mathcal{L}_{03,03} = \frac{4 \kappa}{\hbar \beta} L_0^2 (2 \cos \alpha_1 -\cos^2 \alpha_1 -1) - \frac{1}{2} \mathcal{L}_{11,33}- \frac{1}{2} \mathcal{L}_{22,33},
\end{equation}
\begin{equation}
  \fl \mathcal{L}_{04,04} = \frac{4 \kappa}{\hbar \beta} L_0^2 (-2 \cos \alpha_1 -\cos^2 \alpha_1 -1) - \frac{1}{2} \mathcal{L}_{11,44}- \frac{1}{2} \mathcal{L}_{22,44},
\end{equation}
\begin{equation}
  \fl  \mathcal{L}_{12,12} = -\frac{16 \kappa}{\hbar \beta} L_0^2 \cos^2 \alpha_0 - \frac{1}{2} \mathcal{L}_{00,11}- \frac{1}{2} \mathcal{L}_{00,22},
\end{equation} 
\begin{equation}
  \fl \mathcal{L}_{13,13} = -\frac{4 \kappa}{\hbar \beta} L_0^2 (\cos \alpha_0 - \cos \alpha_1)^2 - \frac{1}{2} \mathcal{L}_{00,11}- \frac{1}{2} \mathcal{L}_{11,33} - \frac{1}{2} \mathcal{L}_{22,33},
\end{equation}
\begin{equation}
 \fl \mathcal{L}_{14,14} = -\frac{4 \kappa}{\hbar \beta} L_0^2 (\cos \alpha_0 + \cos \alpha_1)^2 - \frac{1}{2} \mathcal{L}_{00,11}- \frac{1}{2} \mathcal{L}_{11,44} - \frac{1}{2} \mathcal{L}_{22,44},
\end{equation}
\begin{equation}
  \fl \mathcal{L}_{23,23} = -\frac{4 \kappa}{\hbar \beta} L_0^2 (\cos \alpha_0 + \cos \alpha_1)^2 - \frac{1}{2} \mathcal{L}_{00,22}- \frac{1}{2} \mathcal{L}_{11,33} - \frac{1}{2} \mathcal{L}_{22,33},
\end{equation}
\begin{equation}
  \fl \mathcal{L}_{24,24} = -\frac{4 \kappa}{\hbar \beta} L_0^2 (\cos \alpha_0 - \cos \alpha_1)^2 - \frac{1}{2} \mathcal{L}_{00,22}- \frac{1}{2} \mathcal{L}_{11,44} - \frac{1}{2} \mathcal{L}_{22,44},
\end{equation}
\begin{equation}
 \fl \mathcal{L}_{12,12} = -\frac{16 \kappa}{\hbar \beta} L_0^2 \cos^2 \alpha_1 - \frac{1}{2} \mathcal{L}_{11,33} - \frac{1}{2} \mathcal{L}_{22,33} - \frac{1}{2} \mathcal{L}_{11,44} - \frac{1}{2} \mathcal{L}_{22,44},
\end{equation} 
\begin{eqnarray}
  \fl \mathcal{L}_{01,02} = \frac{4 \kappa}{\hbar \beta} (X_{00}X_{12} - X_{12}X_{22}) - G(\omega_{02}) N_{02} X_{01} X_{02} - G(\omega_{12}) N_{12} X_{11} X_{12} \nonumber\\
   - G(\omega_{32}) N_{32} X_{13} X_{23} - G(\omega_{42}) N_{42} X_{14} X_{24}, 
\end{eqnarray}
\begin{eqnarray}
  \fl \mathcal{L}_{02,01} = \frac{4 \kappa}{\hbar \beta} (X_{00}X_{12} - X_{12}X_{11}) - G(\omega_{01}) N_{01} X_{01} X_{02} - G(\omega_{21}) N_{21} X_{22} X_{12} \nonumber\\
   - G(\omega_{31}) N_{31} X_{13} X_{23} - G(\omega_{41}) N_{41} X_{14} X_{24}, 
\end{eqnarray}
\begin{eqnarray}
  \fl \mathcal{L}_{13,23} = \frac{4 \kappa}{\hbar \beta} (X_{33}X_{12} - X_{12}X_{22}) - G(\omega_{12}) N_{12} (X_{11} X_{12} - X_{12} X_{33}) \nonumber\\
   \fl - G(\omega_{02}) N_{02} X_{01} X_{02} - G(\omega_{32}) N_{32} X_{13} X_{23} - G(\omega_{42}) N_{42} X_{14} X_{24}, 
\end{eqnarray}
\begin{eqnarray}
  \fl \mathcal{L}_{23,13} = \frac{4 \kappa}{\hbar \beta} (X_{33}X_{12} - X_{12}X_{11}) - G(\omega_{21}) N_{21} (X_{22} X_{12} - X_{12} X_{33}) \nonumber\\
   \fl - G(\omega_{01}) N_{01} X_{01} X_{02} - G(\omega_{31}) N_{31} X_{13} X_{23} - G(\omega_{41}) N_{41} X_{14} X_{24}, 
\end{eqnarray}
\begin{eqnarray}
  \fl \mathcal{L}_{14,24} = \frac{4 \kappa}{\hbar \beta} (X_{44}X_{12} - X_{12}X_{22}) - G(\omega_{12}) N_{12} (X_{11} X_{12} - X_{12} X_{44}) \nonumber\\
   \fl - G(\omega_{02}) N_{02} X_{01} X_{02} - G(\omega_{32}) N_{32} X_{13} X_{23} - G(\omega_{42}) N_{42} X_{14} X_{24}, 
\end{eqnarray}
\begin{eqnarray}
  \fl \mathcal{L}_{24,14} = \frac{4 \kappa}{\hbar \beta} (X_{44}X_{12} - X_{12}X_{11}) - G(\omega_{21}) N_{21} (X_{22} X_{12} - X_{12} X_{33}) \nonumber\\
   \fl - G(\omega_{01}) N_{01} X_{01} X_{02} - G(\omega_{31}) N_{31} X_{13} X_{23} - G(\omega_{41}) N_{41} X_{14} X_{24}. 
\end{eqnarray}

\section{Diagonal reduced density matrix elements}\label{AppDiagonalElements}
The solutions of the FSA master equation \eref{SimplifiedMasterEq} for the diagonal elements reads:
\begin{eqnarray} 
  \fl \sigma_{00}(t) = \sigma_{00}^0  + \sigma_{11}^0 + \sigma_{22}^0 + \sigma_{33}^0+ \sigma_{44}^0\nonumber \\
  \fl - \rme^{- \pi \mathcal{L}_{00,11} t} \biggl( \sigma_{11}^0 + \sigma_{33}^0 \frac{\mathcal{L}_{11,33}}{-\mathcal{L}_{00,11} + \mathcal{L}_{11,33} + \mathcal{L}_{22,33}} 
     + \sigma_{44}^0 \frac{\mathcal{L}_{11,44}}{-\mathcal{L}_{00,11} + \mathcal{L}_{11,44} + \mathcal{L}_{22,44}} \biggr) \nonumber \\
  \fl - \rme^{- \pi \mathcal{L}_{00,22} t} \biggl( \sigma_{22}^0 + \sigma_{33}^0 \frac{\mathcal{L}_{22,33}}{-\mathcal{L}_{00,22} + \mathcal{L}_{11,33} + \mathcal{L}_{22,33}}  
                + \sigma_{44}^0 \frac{\mathcal{L}_{22,44}}{-\mathcal{L}_{00,22} + \mathcal{L}_{11,44} + \mathcal{L}_{22,44}} \biggr) \nonumber \\
   \fl + \rme^{- \pi ( \mathcal{L}_{11,33} + \mathcal{L}_{22,33}) t}  \sigma_{33}^0 \biggl(  \frac{ \mathcal{L}_{00,22} - \mathcal{L}_{11,33}  }{-\mathcal{L}_{00,22} + \mathcal{L}_{11,33} + \mathcal{L}_{22,33}} 
            + \frac{ \mathcal{L}_{11,33}  }{-\mathcal{L}_{00,11} + \mathcal{L}_{11,33} + \mathcal{L}_{22,33}}  \biggr) \nonumber\\
   \fl + \rme^{- \pi ( \mathcal{L}_{11,44} + \mathcal{L}_{22,44}) t}  \sigma_{44}^0 \biggl(  \frac{ \mathcal{L}_{00,22} - \mathcal{L}_{11,44}  }{-\mathcal{L}_{00,22} + \mathcal{L}_{11,44} + \mathcal{L}_{22,44}} 
            + \frac{ \mathcal{L}_{11,44}  }{-\mathcal{L}_{00,11} + \mathcal{L}_{11,44} + \mathcal{L}_{22,44}}  \biggr) \label{SolDiagFSA00},
\end{eqnarray}
\begin{eqnarray}
    \fl \sigma_{11}(t) = - \rme^{- \pi \mathcal{L}_{00,11} t}  \sigma_{11}^0 \nonumber  \\
    - \rme^{- \pi ( \mathcal{L}_{00,11} + \mathcal{L}_{11,33} + \mathcal{L}_{22,33}) t}  \sigma_{33}^0  \frac{ \mathcal{L}_{11,33}  }{-\mathcal{L}_{00,11} + \mathcal{L}_{11,33} + \mathcal{L}_{22,33}} \nonumber\\
     - \rme^{- \pi (  \mathcal{L}_{00,11} +\mathcal{L}_{11,44} + \mathcal{L}_{22,44}) t}  \sigma_{44}^0  \frac{ \mathcal{L}_{11,44}  }{-\mathcal{L}_{00,11} + \mathcal{L}_{11,44} + \mathcal{L}_{22,44}}, \label{SolDiagFRWA11}
\end{eqnarray}
\begin{eqnarray}
  \fl \sigma_{22}(t) = - \rme^{- \pi \mathcal{L}_{00,22} t}  \sigma_{22}^0 \nonumber \\
    - \rme^{- \pi ( \mathcal{L}_{00,22} + \mathcal{L}_{11,33} + \mathcal{L}_{22,33}) t}  \sigma_{33}^0  \frac{ \mathcal{L}_{22,33}  }{-\mathcal{L}_{00,22} + \mathcal{L}_{11,33} + \mathcal{L}_{22,33}} \nonumber\\
    - \rme^{- \pi ( \mathcal{L}_{00,22} + \mathcal{L}_{11,44} + \mathcal{L}_{22,44}) t}  \sigma_{44}^0  \frac{ \mathcal{L}_{22,44}  }{-\mathcal{L}_{00,22} + \mathcal{L}_{11,44} + \mathcal{L}_{22,44}}, \label{SolDiagFRWA22}
\end{eqnarray}
\begin{equation}
  \fl \sigma_{33}(t) =  \rme^{- \pi ( \mathcal{L}_{11,33} + \mathcal{L}_{22,33}) t}  \sigma_{33}^0, \label{SolDiagFRWA33}
\end{equation} 
\begin{equation}
  \fl \sigma_{44}(t) =  \rme^{- \pi ( \mathcal{L}_{11,44} + \mathcal{L}_{22,44}) t}  \sigma_{44}^0. \label{SolDiagFSA44}
\end{equation}

\end{document}